\documentclass[12pt]{iopart}

\expandafter\let\csname equation*\endcsname\relax
  \expandafter\let\csname endequation*\endcsname\relax
\usepackage[usenames,dvipsnames]{color} 
\usepackage{graphicx,amssymb,amsmath}
\usepackage{dcolumn}

\usepackage{multirow}
\usepackage{doi}
\renewcommand{\emph}{\textit}
\usepackage{hyperref}
\usepackage{CJK}
\usepackage{subcaption}

\hypersetup{colorlinks, linkcolor=blue,anchorcolor=blue,citecolor=blue,urlcolor=blue}

\newcommand{\ket}[1]{\left\vert{#1}\right\rangle}

\begin{document}
\begin{CJK*}{UTF8}{fs}
\title{Coherence protection and decay mechanism in qubit ensembles under concatenated continuous driving}
\author{Guoqing Wang \CJKfamily{gbsn}(王国庆)}
\address{
   Research Laboratory of Electronics and Department of Nuclear Science and Engineering, Massachusetts Institute of Technology, Cambridge, MA 02139}
\author{Yi-Xiang Liu \CJKfamily{gbsn}(刘仪襄)}
\address{
   Research Laboratory of Electronics and Department of Nuclear Science and Engineering, Massachusetts Institute of Technology, Cambridge, MA 02139}
\author{Paola Cappellaro}
\address{
   Research Laboratory of Electronics and Department of Nuclear Science and Engineering, Massachusetts Institute of Technology, Cambridge, MA 02139}
\address{Department of Physics, Massachusetts Institute of Technology, Cambridge, MA 02139 }
\ead{pcappell@mit.edu}

\begin{abstract}

Dense ensembles of spin qubits are valuable for quantum applications, even though their coherence protection remains challenging. 
Continuous dynamical decoupling can protect ensemble qubits from noise while allowing gate operations, but it is hindered by the additional noise introduced by the driving. Concatenated continuous driving (CCD) techniques can, in principle, mitigate this problem. Here we provide deeper insights into the dynamics under CCD, based on Floquet theory, that lead to optimized state protection by adjusting driving parameters in the CCD scheme to induce mode evolution control. We experimentally demonstrate the improved control by simultaneously addressing a dense Nitrogen-vacancy (NV) ensemble with $10^{10}$ spins. We achieve an experimental 15-fold improvement in coherence time for an arbitrary, unknown state, and a 500-fold improvement for an arbitrary, known state, corresponding to driving the sidebands and the center band of the resulting Mollow triplet, respectively. 
We can achieve such coherence time gains by optimizing the driving parameters to take into account the noise affecting our system.  By extending the generalized Bloch equation approach to the CCD scenario, we identify the noise sources that dominate the decay mechanisms in NV ensembles, confirm our model by experimental results, and identify the driving strengths yielding optimal coherence. 
Our results can be directly used to optimize qubit coherence protection under continuous driving and bath driving, and enable applications in robust pulse design and quantum sensing.
\end{abstract}

\maketitle
\end{CJK*}

\section{Introduction}

Scaling up the size of quantum systems is desirable in many quantum technologies, ranging from quantum simulators to quantum sensors. 
However, manipulating a large quantum system while simultaneously protecting the coherence remains challenging, even when the quantum application only requires collective control, such as some special ensemble-based quantum sensors or simulators. In particular, frequency and driving inhomogeneities typically increase when the system size increases. Various techniques such as pulsed and continuous dynamical decoupling \cite{hahnSpinEchoes1950,carrEffectsDiffusionFree1954,meiboomModifiedSpinEcho1958,ryanRobustDecouplingTechniques2010a,souzaRobustDynamicalDecoupling2012,uhrigKeepingQuantumBit2007,yangUniversalityUhrigDynamical2008,biercukOptimizedDynamicalDecoupling2009,mukhtarUniversalDynamicalDecoupling2010,violaRobustDynamicalDecoupling2003,gordonOptimalDynamicalDecoherence2008,fanchiniContinuouslyDecouplingSinglequbit2007,hiroseContinuousDynamicalDecoupling2012,laraouiRotatingFrameSpin2011,mkhitaryanDecayRotaryEchoes2014,aharonFullyRobustQubit2016,starkClockTransitionContinuous2018}, as well as spin-locking \cite{loretzRadioFrequencyMagnetometryUsing2013} have been used to protect the coherence of quantum systems. 

Beyond achieving robust quantum memories, manipulating the quantum device while maintaining its coherence remains a non-trivial task~\cite{DuCoherenceProtectedPhysRevLett.109.070502}, but could be helped by using continuous decoupling schemes. Unfortunately, these often introduce additional sources of noise linked to the added driving fields.
A technique termed concatenated continuous driving (CCD), which consists of adding multiple resonant modulated fields,  can combat external noise and fluctuations in the control fields \cite{caiRobustDynamicalDecoupling2012,khanejaUltraBroadbandNMR2016,saikoSuppressionElectronSpin2018,cohenContinuousDynamicalDecoupling2017,farfurnikExperimentalRealizationTimedependent2017,rohrSynchronizingDynamicsSingle2014,laytonRabiResonanceSpin2014,saikoMultiphotonTransitionsRabi2015,teissierHybridContinuousDynamical2017,bertainaExperimentalProtectionQubit2020,caoProtectingQuantumSpin2020}. 
A modulation field on resonance with the main driving term can suppress decoherence, provided that its amplitude is much larger than the fluctuations of the main driving. 

Here, we use the CCD scheme to protect the coherence of an ensemble of qubits and to achieve their collective manipulation in the presence of frequency and driving field inhomogeneities. 
Experimentally, we achieve a 15-fold improvement  in the coherence time for an arbitrary, unknown state (corresponding to the transverse coherence time) and we also show how to tune the CCD control to protect an arbitrary, known state  with  a 500-fold improvement in its coherence. 
These results are achieved through a more comprehensive understanding of the modulated dynamics, which can be described by Floquet theory as giving rise to a Mollow triplet~\cite{wang2020observation}.
A strong modulation has been demonstrated to have a broad feature in the synchronization by evaluating the power and detuning dependence of the evolution amplitude. 

The long coherence times we achieve are also predicated on selecting the optimal control parameters given the characteristics of the noise. We thus carefully characterize the experimental noise sources by evaluating the power and detuning dependences of the Rabi signal coherence, and analyze the noise effects under the CCD scheme by extending the theoretical framework of the generalized Bloch equation to this scenario. This analysis, confirmed by experimental results, allows not only to optimize the coherence time by adjusting the drive parameters, but it could be also used to reconstruct the power spectral density (PSD) of various noise sources. Finally, we briefly discuss the potential applications in the protection of nuclear spin coherence, perfect pulse design and AC magnetic field sensing.

\begin{figure}[htbp]
\centering \includegraphics[width=160mm]{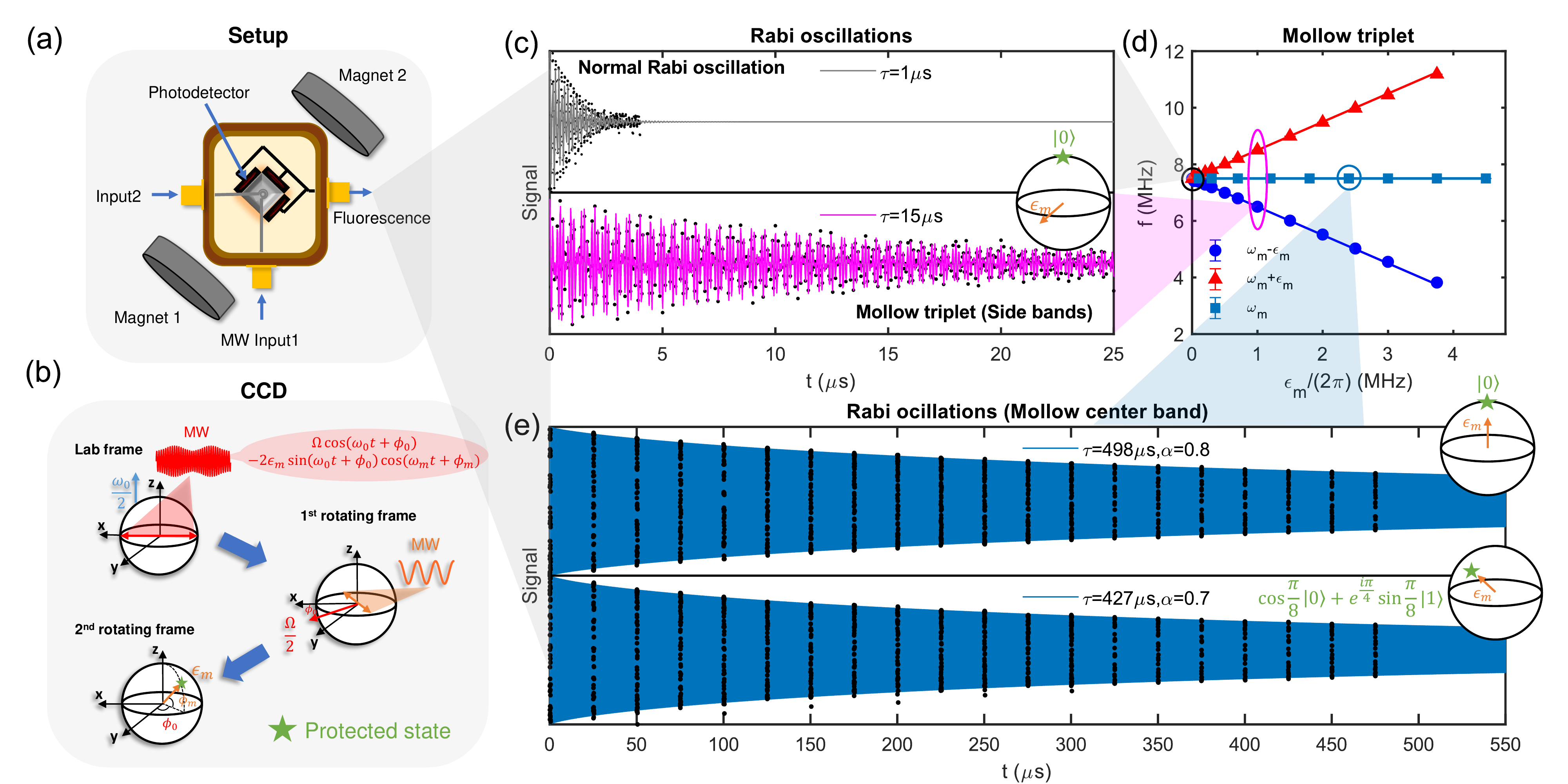}
\caption{\label{LongCoherence} CCD for coherence protection. (a) Schematic of the experimental setup. (b) Principle of the amplitude-modulated CCD scheme. Upon applying a modulated waveform, we can enter into two rotating frames where direction and strength of the driving field can be set by the modulation parameters $\epsilon_m,\phi_0,\phi_m$. (c) Coherence protection of the Mollow triplet sidebands with the phase-modulated CCD scheme. Parameters $\Omega=\omega_m=(2\pi)7.5\text{MHz},\phi_m=0$ are chosen such that the effective driving field in the second rotating frame is perpendicular to the initial state $|0\rangle$. Upper panel is a normal Rabi oscillation with $\epsilon_m=0$. Lower panel is with $\epsilon_m=(2\pi)1\text{MHz}$. Coherence times $\tau_i$ are fitted with $c_0+\sum_{i} c_ie^{-\frac{t}{\tau_i}}\cos(\omega_it+\phi_i)$. (d) Mollow triplet in the CCD scheme. Parameters $\Omega=\omega_m=(2\pi)7.5\text{MHz}$, and initial state is prepared to $\ket{0}$. Sidebands are measured with $\phi_m=0$ and center band is measured with $\phi_m=\pi/2$. Frequency values are fitted from the Rabi oscillations with the same function as in (c). Solid lines are the theoretical predictions of the frequency values $\omega_m$, $\omega_m\pm\epsilon_m$. (e) Coherence protection of the center band. Modulation strength $\epsilon_m=(2\pi)2.4\text{MHz}$. The driving field in the second rotating frame is adjusted to the same direction as the initial state such that the evolution will only involve the center band. In upper panel, the initial state is $|0\rangle$ and the driving parameters are $\phi_0=0,\phi_m=\pi/2$. In lower panel, the initial state is $\cos(\frac{\pi}{8})|0\rangle+e^{i\frac{\pi}{4}}\sin(\frac{\pi}{8})|1\rangle$ and the driving parameters are $\phi_0=-\frac{\pi}{4},\phi_m=\frac{\pi}{4}$. Coherence time $\tau$ and index $\alpha$ are fitted with $c_0+c_1e^{-(\frac{t}{\tau})^\alpha}\cos(\omega_1t+\phi_1)$.}
\end{figure}

\section{Coherence protection with the CCD scheme}
\subsection{Setup}
NV centers in diamond have emerged as a promising platform for quantum information processing \cite{dohertyNitrogenvacancyColourCentre2013a}, thanks in part to good control techniques that have pushed their coherence times nearly up to  the relaxation limit \cite{phamEnhancedSolidstateMultispin2012,bauchUltralongDephasingTimes2018,naydenovDynamicalDecouplingSingleelectron2011,shimRobustDynamicalDecoupling2012,bar-gillSolidstateElectronicSpin2013,caiRobustDynamicalDecoupling2012,farfurnikExperimentalRealizationTimedependent2017,teissierHybridContinuousDynamical2017,caoProtectingQuantumSpin2020}. 
Our device is based on an ensemble of NV centers in diamond as previously reported in Ref.~\cite{jaskulaPhysRevApplied.11.054010} [see Fig.~\ref{LongCoherence}(a) for a schematic of the setup]. 
A pair of permanent magnets apply a static magnetic field along the NV axis,  $B_0\approx 230$G, which lifts the degeneracy of the $|m_S=\pm1\rangle$ states. The energy gap between the $|m_S=0\rangle$ and $|m_S=-1\rangle$ states that we address in experiments is 2.207GHz when the $^{14}\text{N}$ nuclear spin is in state $|m_I=1\rangle$.  
Laser illumination not only initializes the NV electronic spin in the $\ket{m_S=0}$ state, but also polarizes the $^{14}\text{N}$ nuclear spin states to 73\% in $|m_I=1\rangle$. 
Microwave is delivered through a 0.7mm loop structure on a PCB board. Three photodiodes are attached to the surface of the diamond, and glued on the same PCB to measure the fluorescence. 
By focusing a 0.4mW green laser beam to a $30\mu m$ spot, we simultaneously address $\sim 10^{10}$ spins.
An arbitrary waveform generator mixes a $\sim 100$MHz frequency with a carrier microwave frequency generated by a signal generator to implement the coherent control. By applying a resonant microwave, we selectively address NV electronic spin $|m_S=0\rangle$ and $|m_S=-1\rangle$ as the logical $|0\rangle$ and $|1\rangle$ states of an effective qubit.

\subsection{Coherence protection}
Due to field and driving inhomogeneities across the sample, the coherence time under normal Rabi driving is about $1\mu$s [see Fig.~\ref{LongCoherence}(c) upper panel]. To overcome these limitations, we use a CCD scheme, whose basic principles are shown in Fig.~\ref{LongCoherence}(b). 
Consider a two-level system with a static splitting $\omega_0$ along z, coupled to an amplitude-modulated microwave along the x axis $\Omega\cos(\omega t+\phi_0)-2\epsilon_m\sin(\omega t+\phi_0)\cos(\omega_m t+\phi_m)$. 
When the rotating wave approximation (RWA) condition $\Omega,\epsilon_m\ll\omega_0$ is satisfied and $\phi_0=0$, going into the first rotating frame defined by $H_0=\frac{\omega}{2}\sigma_z$ and neglecting the counter-rotating term, the Hamiltonian becomes
\begin{equation}
    H_I=-\frac{\delta}{2}\sigma_z+\frac{\Omega}{2}\sigma_x+\epsilon_m\cos(\omega_mt+\phi_m)\sigma_y
    \label{HI_AmpMod}
\end{equation} 
Phase modulation can also engineer a similar Hamiltonian through a phase-modulated waveform $\Omega\cos\left[\omega t+\frac{2\epsilon_m}{\Omega}\cos(\omega_m t+\phi_m)\right]$. In the first rotating frame defined by $H_0(t)=\frac{\omega}{2}\sigma_z- \frac{\epsilon_m\omega_m}{\Omega}\sin(\omega_m t+\phi_m)\sigma_z$, the Hamiltonian becomes 
\begin{equation}
H_I=-\frac{\delta}{2}\sigma_z+\frac{\Omega}{2}\sigma_x+\epsilon_m\frac{\omega_m}{\Omega}\sin(\omega_mt+\phi_m)\sigma_z.
\label{HI_FreqMod}
\end{equation}
When the second RWA condition $\epsilon_m\ll\Omega$ is satisfied, the Rabi oscillations display contributions from a center band $\omega_m$ and two sidebands $\omega_m\pm\sqrt{\epsilon_m^2+(\omega_m-\Omega_R)^2}$,  forming the Mollow triplet [Fig.~\ref{LongCoherence}(d)]. 
The intensity of the center  and sidebands can be tuned by the phases $\phi_0,\phi_m$ of the driving (\textit{mode control}). 
When the initial state is in the direction of the driving field in the second rotating frame, only the center band appears; when the initial state is perpendicular to the field, only the sidebands exist. 
For a generic initial state, all three bands contribute to the signal. 
Beyond the RWA, higher order frequency components as well as frequency and amplitude shifts complicate the dynamics, but nevertheless their effects can be precisely predicted by  Floquet theory~\cite{wang2020observation}. Here we focus on the dynamics within the RWA.

The two sidebands are affected by fluctuations of the second driving field, whereas the center band frequency is robust against noise, as it depends only on the modulation frequency which is set with high precision. Thus, while a generic, \textit{unknown} state coherence is limited by the shorter, sideband coherence, we can use our knowledge of the central band dynamics to better protect a \textit{known}, arbitrary state,  by synchronizing the  mode of the center band to the qubit state. 
To demonstrate these improvements, in experiments we evaluate the coherence improvement of the center band and sidebands separately, by setting different modulation phases and initial states. 
First, we show in Fig.~\ref{LongCoherence}(d) that the coherence of the sidebands displays a large improvement by more than an order of magnitude, when compared to a normal Rabi oscillation. 
By further tuning the parameters $\phi_0,\phi_m$ in the CCD scheme, we can orient the driving field in the second rotating frame along the direction of the initial state to be protected. This synchronizes the state evolution  to the Mollow center band, and achieves a  $500$-fold improvement in the coherence time, compared to the conventional Rabi oscillations.
Fig.~\ref{LongCoherence}(e) shows the coherence of  two different initial states synchronized to the center band. In the upper panel, the initial state is $|0\rangle$ and the driving phases are $\phi_0=0,\phi_m=\pi/2$; in the lower panel, the initial state is $\cos(\frac{\pi}{8})|0\rangle+e^{i\frac{\pi}{4}}\sin(\frac{\pi}{8})|1\rangle$ and the driving phases are $\phi_0=-\frac{\pi}{4},\phi_m=\frac{\pi}{4}$. Note that the coherence times of both states are similar, indicating that an arbitrary known state can be protected.  

\begin{figure}[htbp]
\centering \includegraphics[width=130mm]{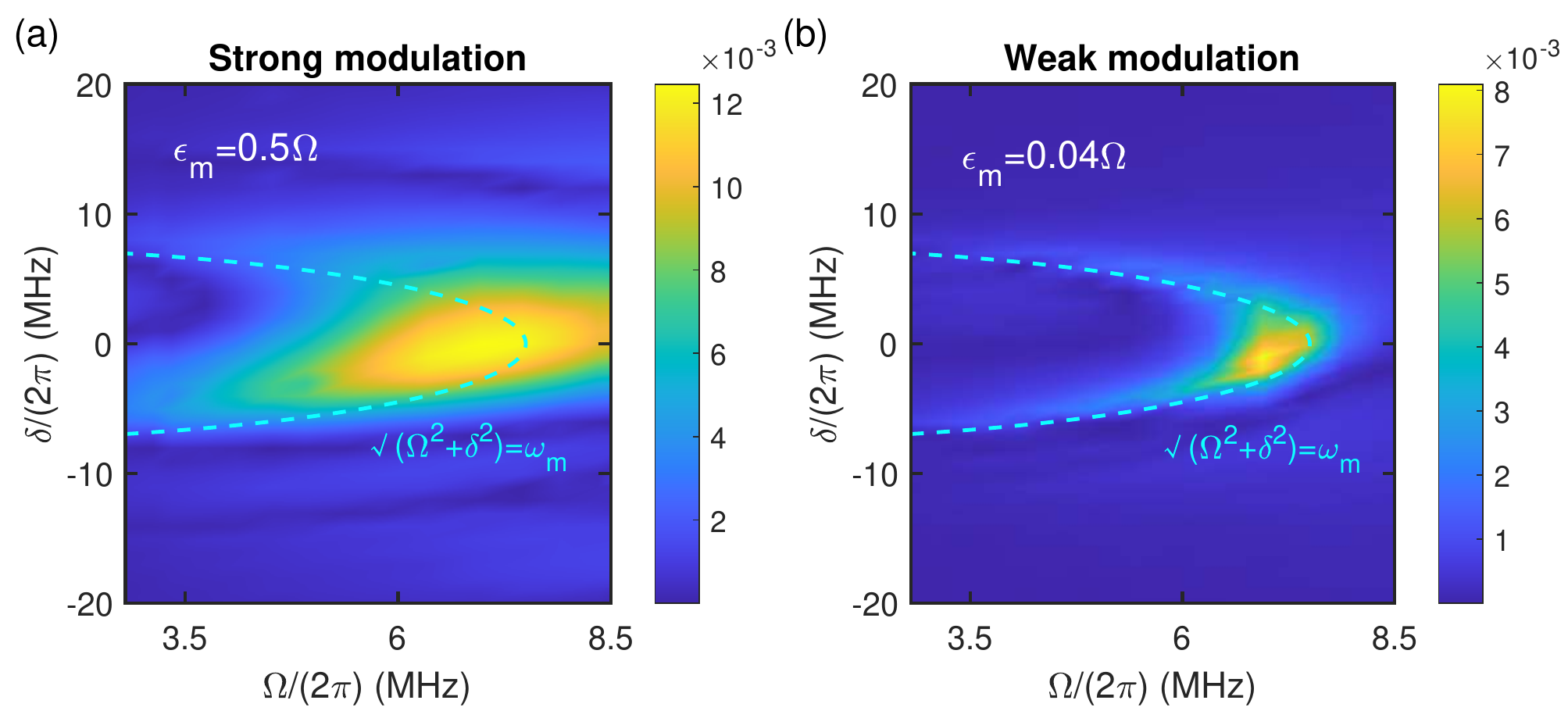}
\caption{
\label{2DRobustness} 
Synchronization of a single mode evolution. (a) Power and detuning dependence of the center band oscillation amplitude. Initial state is prepared to $\ket{0}$ and modulation frequency is $\omega_m=(2\pi)7.5\text{MHz}$. An amplitude-modulated CCD waveform is applied with a strong modulation condition $\epsilon_m=\frac{1}{2}\Omega$ kept unchanged. $\phi_m=\pi/2$ such that the evolution is synchronized to the center band when $\Omega=\omega_m$, $\delta=0$. Rabi oscillations are measured from $50\mu$s to $50.5\mu$s to ensure that only the center band is alive. The intensity represents the value of contrast $c_1$ fitted from Rabi oscillations with $c_0+\frac{1}{2}c_1\cos(\omega_1 t+\phi_1)$. (b) Similar experiment with a weak modulation $\epsilon_m=\frac{1}{25}\Omega, \phi_m=\pi/2$. Note that the signal contrast of a normal Rabi oscillation measured in our sample is around 1\%-2\%.}
\end{figure}

We further study the robustness of the mode-synchronized driving protocol against inhomogeneities in the driving and static fields that occur when manipulating large ensemble of spins. 
We measure the Rabi oscillations from $t=50\mu $s to $t=50.5\mu $s to ensure that only the center band survives, and extract the oscillation contrast $c_1$ by fitting the signal to $c_0+\frac{1}{2}c_1\cos(\omega_1 t+\phi_1)$. In Fig.~\ref{2DRobustness}, we compare the results with a (a) strong and (b) weak modulation strength $\epsilon_m$. In the first case, the center band has a large amplitude in a broader region beyond the resonance condition $\sqrt{\Omega^2+\delta^2}=\omega_m$, showing that more spins are driven even if their detuning and Rabi frequency deviates from the nominal ones due to inhomogeneities. 
Another evidence of robustness is that the measured oscillation contrast (intensity in the color map) at a nominal $\delta=0,\Omega=\omega_m$ under strong modulation is larger than that under weak modulation, indicating that the strong modulation improves the protection of the center band coherence. 

We note that similar order-of-magnitude improvements in the qubit coherence had only been observed for single NV centers ~\cite{caiRobustDynamicalDecoupling2012}, or for small ensembles of NVs~\cite{farfurnikExperimentalRealizationTimedependent2017,caoProtectingQuantumSpin2020}, while here we are able to engineer robust control over a large volume consisting of an ensemble of $10^{10}$ NV spins. In addition, we identified the mechanism for robust protection of known quantum states via mode control, that was only previously achieved with mechanical driving~\cite{teissierHybridContinuousDynamical2017}.

\section{Coherence time analysis}
To further understand the protection afforded by the CCD scheme, as well as select the optimal driving parameters, it is critical to develop a theoretical framework for the coherence time of qubit ensembles under this scenario, and implement experiments to  verify the theoretical predictions.

In the regime of a qubit weakly coupled to the bath, its decay rate under a single transverse driving field can be predicted by the generalized Bloch equation (GBE) where the relaxation rates are given by the spectral components of the noise on resonance with the corresponding transition energies of the qubit~\cite{gevaRelaxationTwoLevel1995}. 
The coherence time of the qubit is thus determined by the power spectral density (PSD) of the noise \cite{jingDecoherenceElectricallyDriven2014,yanRotatingframeRelaxationNoise2013}. Here we generalize the GBE model to an ensemble of spins,  modeling the ensemble as a single spin qubit, where  field inhomogeneities are included as an additional zero-frequency component in the noise spectrum. 
With a semi-classical treatment, the field fluctuations can be included as a stochastic component in the amplitude-modulated CCD Hamiltonian, yielding
\begin{equation}
    H=\frac{\omega_0}{2}\sigma_z+(\Omega+\xi_{\Omega})\cos(\omega t)\sigma_x-2(\epsilon_m+\xi_{\epsilon_m})\sin(\omega t)\cos(\omega_m t+\phi_m))\sigma_x+\xi_x\sigma_x+\xi_z\sigma_z
\end{equation} 
where $\xi_{\Omega}$, $\xi_{\epsilon_m}$ are the fluctuations of the driving fields, comprising both  driving fluctuations and  inhomogeneities. $\xi_x$, $\xi_z$ are the fluctuations of the transverse and longitudinal field giving rise to $T_1$ and $T_2$ decay in the absence of driving, with contributions from both the bath and the static field inhomogeneities. 
Assuming stationary processes, the time correlation of these fluctuations  is the Fourier transformation of their noise PSD,  $\langle\xi_j(t_1)\xi_j(t_2)\rangle=\frac{1}{2\pi}\int_{-\infty}^{\infty} d\nu S_j(\nu)e^{-i\nu (t_2-t_1)}$ where $S_j(\nu)$ $(j=x,z,\Omega,\epsilon_m)$ is the PSD of the corresponding noise in the lab frame. 
In the same way we can better understand the unitary dynamics by  applying rotating frame transformations to the Hamiltonian, here we can analyze the noise effects and derive the expected decay rates  by  expressing the PSDs in the rotating frame as a function of the PSDs in the lab frame \cite{gevaRelaxationTwoLevel1995,yanRotatingframeRelaxationNoise2013,jingDecoherenceElectricallyDriven2014}. 
This is important as only some frequency components of the PSD contribute mostly to the decay in any given frame: the transverse on-resonance noise component contributes to the qubit random bit flips, whereas the longitudinal noise components at zero frequency contribute to random phase flips. 

\begin{figure}[htbp]
\centering \includegraphics[width=130mm]{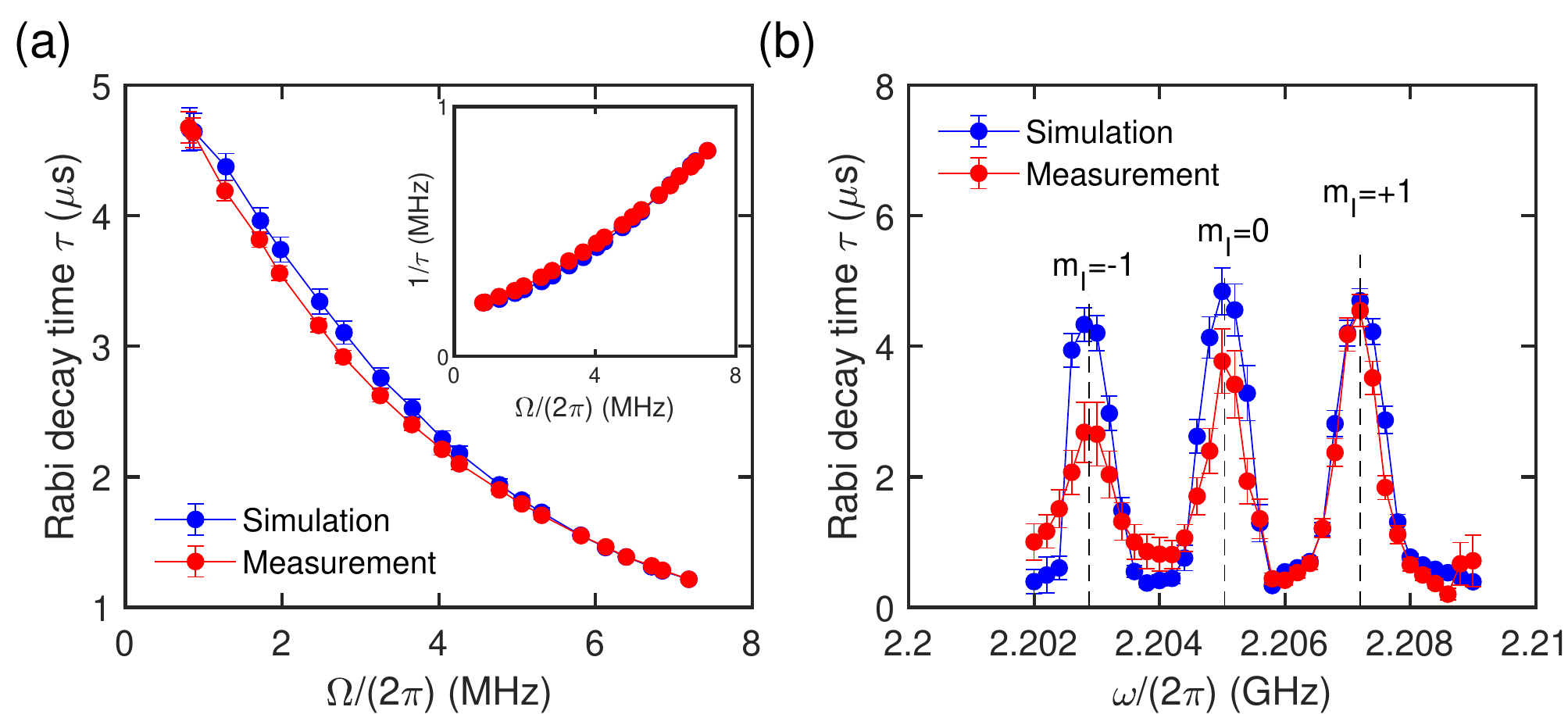}
\caption{\label{Noise}  Inhomogeneity characterization. (a) Power dependence of the Rabi coherence time $T_{2\rho}$. Microwave frequency $\omega=\omega_0=(2\pi)2.2072$GHz is on resonance with the $|m_I=+1\rangle$ sublevel of nuclear spin of $^{14}\text{N}$. (b) Detuning dependence of the Rabi coherence time $T_{2\rho}$. Microwave power is chosen such that Rabi frequency under resonance condition is $0.7$MHz. Simulations plotted in blue points are calculated by integrating the Rabi oscillation over power and detuning distributions $f(\Omega+\xi_\Omega,\omega+\xi_z)=\frac{1}{2\pi \sigma_\Omega \sigma_{\omega}}\exp(-\frac{\xi_\Omega^2}{2\sigma_\Omega^2}-\frac{\xi_\omega^2}{2\sigma_{\omega}^2})$ and summing up the three species of nuclear spin sublevels with the population of each sublevel obtained from the ESR measurement. Values $\sigma_\Omega=0.016\Omega$ and $\sigma_{\omega}=(2\pi)0.32\text{MHz}$ are used in the simulation. An exponential decay $c_0+c_1\exp(-\frac{t}{\tau})\cos(\omega_1 t+\phi_1)$ is used in the fitting to extract the coherence time $\tau$.}
\end{figure}

For a single driving field ($\epsilon_m=\xi_{\epsilon_m}=0$) and under the resonance condition $\omega=\omega_0$, the longitudinal and transverse relaxation times in the first rotating frame, $T_{1\rho}$, $T_{2\rho}$, are \begin{align}
    \frac{1}{T_{1\rho}}&=\frac{1}{2} S_x(\omega_0)+S_z(\Omega)=\frac{1}{2T_1}+S_z(\Omega)\\
    \frac{1}{T_{2\rho}}&=\frac{1}{2T_{1\rho}}+\frac{1}{2}S_x(\omega_0)+\frac{1}{4}S_{\Omega}(0).
\end{align}
Given long $T_1$ relaxation times, the longitudinal relaxation 
$T_{1\rho}$, corresponding to the spin-locking condition, is dominated  by $S_z(\Omega)$, the longitudinal field fluctuations. 
For a zero-frequency centered noise spectrum, larger driving strengths $\Omega$ result in better coherence as $S_z(\Omega)$ picks the noise at a frequency farther away from zero. 
The transverse relaxation time $T_{2\rho}$ describes the decay of a conventional Rabi oscillation. 
The dominant terms are typically $\frac{1}{2}S_z(\Omega)+\frac{1}{4}S_\Omega(0)$, leading to competing effects as a function of $\Omega$. When $\Omega$ is increased, $\frac{1}{2}S_z(\Omega)$ decreases but $\frac{1}{4}S_\Omega(0)$ increases. 
In Fig.~\ref{Noise}(a), we study the driving strength dependence of the Rabi coherence. Since the coherence time monotonically decreases in the measured range, $\frac{1}{4}S_\Omega(0)$ is the dominant source. 
When the Rabi driving is off-resonance, the Hamiltonian in the rotating frame has components in the $x-z$ plane. Then, the transverse decay rate includes a term $\propto \frac{\delta^2}{\Omega^2+\delta^2}S_z(0)$ [see details in~\ref{Appendix_CoherenceLimit}] that soon dominates since it probes the spectrum at zero frequency. 
Thus, the Rabi coherence  dependence on the detuning $\delta$ provides  information about the static field fluctuation $S_z(0)$, and locally optimal coherence is obtained under three resonance frequencies corresponding to three nuclear spin sub-levels. 
To extract the values of inhomogeneities from the experimental data, we simulate the decay rate with a simple model by directly integrating the Rabi oscillation over a static Gaussian distribution of the driving field inhomogeneities  $\xi_\Omega$ and static field inhomogeneities $\xi_z$. The optimal fit to experiments is obtained with parameters $\sigma_\Omega\sim 1.6\%\Omega$ and  $\sigma_{\omega}=4S_z(0)=(2\pi)0.32\text{MHz}$, or $\sigma_{\omega}\approx0.015\%\omega$.

\begin{figure}[htbp]
\centering \includegraphics[width=150mm]{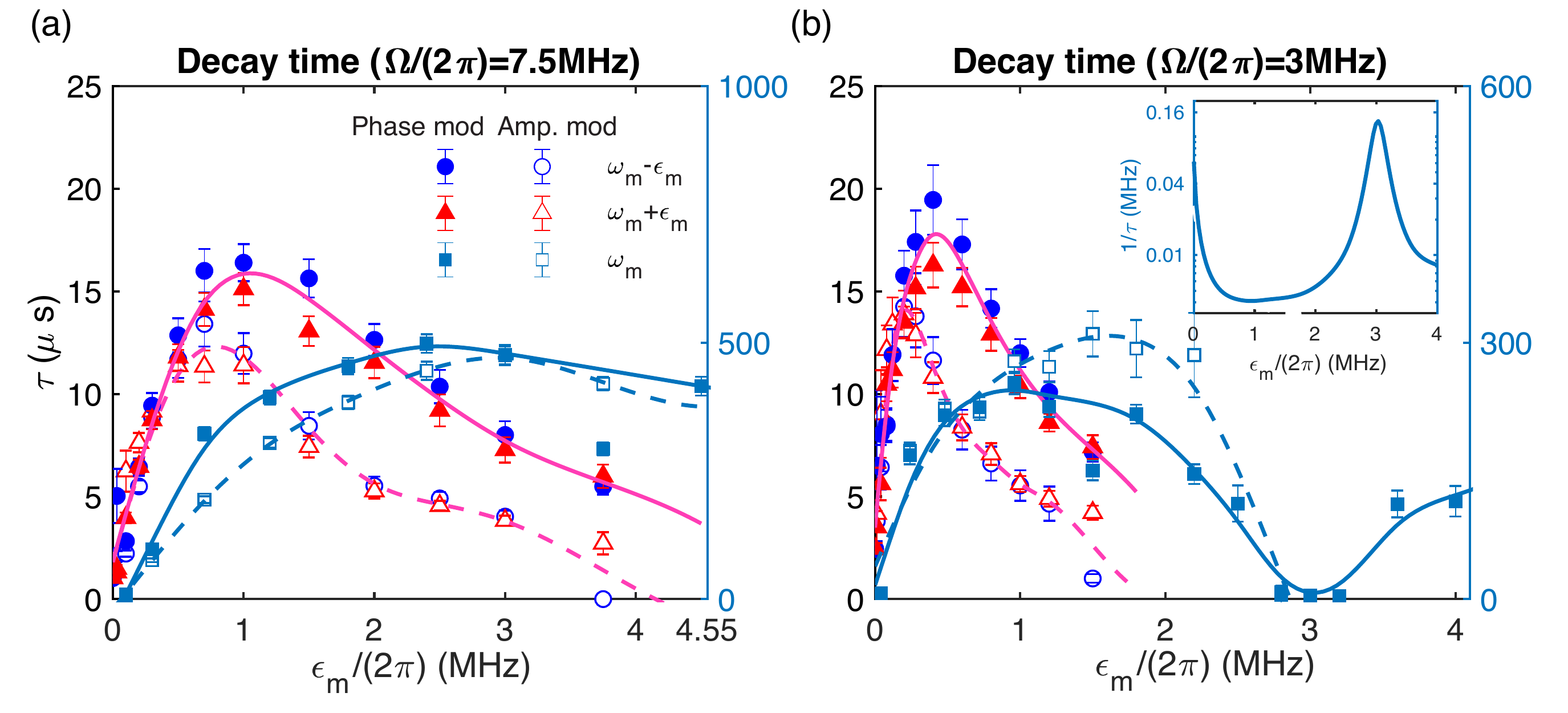}
\caption{\label{DecaytimeVSem} (a)  Coherence times $T_{1\rho\rho}$, $T_{2\rho\rho}$ as a function of $\epsilon_m$. Parameters $\omega_m=\Omega=(2\pi)7.5\text{MHz},\delta=0$, and initial state is prepared to $\ket{0}$. Sidebands coherence times are plotted in red points for $\omega_m+\epsilon_m$ and blue points for $\omega_m-\epsilon_m$. 
Center band coherence times are plotted in light blue points. Solid points and curves are for phase modulation whereas hollow points and dashed lines are for amplitude modulation. Two $y$ axes are used for sidebands (left) and center band (right). (b) Coherence times $T_{1\rho\rho}$,$T_{2\rho\rho}$ dependence on $\epsilon_m$. Parameters $\omega_m=\Omega=(2\pi)3\text{MHz}$. The inset is the decay rate $1/\tau$ of the center band under the phase modulation condition.  }
\end{figure}

We can extend this analysis to the CCD protocol by entering a second rotating frame. On resonance $\omega_m=\Omega$, we obtain the longitudinal and transverse relaxation times $T_{1\rho\rho},T_{2\rho\rho}$ in the second rotating frame, 
\begin{align}
    \frac{1}{T_{1\rho\rho}}&=\frac{1}{4}S_{\Omega}(\epsilon_m)+\frac{3}{4}S_x(\omega_0)  \label{T1rr}+\frac{1}{4}[S_z(\Omega-\epsilon_m)+S_z(\Omega+\epsilon_m)]\\
    \frac{1}{T_{2\rho\rho}}&=\frac{1}{2T_{1\rho\rho}}+\frac{1}{4} S_{\epsilon_m}(0)+\frac{1}{2}S_z(\Omega)+\frac{1}{4}S_x(\omega_0) \label{T2rr}\\
    &=\frac{1}{4} S_{\epsilon_m}(0)+\frac{1}{8}S_\Omega(\epsilon_m)+\frac{1}{2}S_z(\Omega)+\frac{1}{8}[S_z(\Omega-\epsilon_m)+S_z(\Omega+\epsilon_m)]+\frac{5}{8}S_x(\omega_0)\nonumber
\end{align}
$T_{1\rho\rho}$ is the coherence time under the spin-locking condition in the second rotating frame, which corresponds to the center band in the Mollow triplet. $T_{2\rho\rho}$ is the coherence time for the sidebands. By analyzing the dominant noise sources, we can explain the decay rates observed in experiments, as shown in Fig.~\ref{DecaytimeVSem}, and propose good control strategies.

The coherence time $T_{1\rho\rho}$ shows a strong dependence on the driving powers, $\epsilon_m$. 
The fast initial increase in coherence time is due to the fast decrease of $S_{\Omega}(\epsilon_m)$ as $\epsilon_m$ grows, followed by a broad plateau. When $\epsilon_m$ approaches $\Omega$, a fast decrease happens in the coherence time as observed in Fig.~\ref{DecaytimeVSem}(b) due to the  increase of the noise term  $\frac{1}{4}S_{z}(\Omega-\epsilon_m)$ around zero frequency. 
The values of the optimal coherence time $T_{1\rho\rho}$ in both (a) and (b) approach the spin-locking coherence $T_{1\rho}$ under the same driving strength $\Omega$ [see details in Fig.~\ref{SpinLocking} in~\ref{Appendix_CoherenceLimit}], which verifies that the coherence of both the spin-locking condition and the center band in the CCD scheme is dominated by $S_z$. This result points to a strategy to improve the spin-locking coherence time under CDD, by increasing the second drive strength past the first, $\epsilon_m>\Omega$. We note that in this regime, high-order Floquet effects need to be taken into account~\cite{wang2020observation}. 
Since no $S_{\epsilon_m}$ term is involved in the center band coherence, phase modulation and amplitude modulation do not display a significant difference.

The coherence time for the sidebands $T_{2\rho\rho}$ also shows a maximum as a function of $\epsilon_m$. This is due to the competing effects of  $\frac{1}{8}S_\Omega(\epsilon_m)+\frac{1}{8}S_z(\Omega+\epsilon_m)$, which  decreases with increasing $\epsilon_m$ and $\frac{1}{4}S_{\epsilon_m}(0)+\frac{1}{8}S_z(\Omega-\epsilon_m)$, which instead increases. 
Since $\frac{1}{4}S_{\epsilon_m}(0)$ always picks up the DC noise components, it soon dominates when $\epsilon_m$ keeps increasing,  so that the coherence degradation  happens earlier than for the center band. Characterizing the various noise spectrum components can inform the best driving parameters for optimal coherence. 

For an ideal situation of phase modulation, we should have  $\xi_{\epsilon_m}=0$, $\frac{1}{4}S_{\epsilon_m}(0)=0$, and the decrease of coherence time $T_{2\rho\rho}$ should only be induced by the $\frac{1}{8}S_z(\Omega-\epsilon_m)$ term, predicting a coherence time close to $T_{1\rho\rho}$. 
However, both previous experiments \cite{farfurnikExperimentalRealizationTimedependent2017} and this work do not find significant difference between phase modulation and amplitude modulation, although indeed the maximum is shifted towards higher $\epsilon_m$. 
In both strong power and weak power cases in Figs.~\ref{DecaytimeVSem}(a) and \ref{DecaytimeVSem}(b) respectively, the phase modulation only shows a slight improvement for the sidebands and the coherence time of the sidebands is 1-2 orders of magnitude smaller than the center band. The noise $\xi_{\epsilon_m}$ under the phase modulation may come from the phase noise of the microwave field.

\section{Conclusion}
In this work, we explore optimal coherence protection by the CCD technique in dense NV ensembles. We show that any arbitrary states can be protected by aligning the driving field with the state to be protected, thus engineering a single mode evolution that corresponds to the center band in the Mollow triplet. Our experiments show that such a technique can be used to synchronize the dynamics of qubit ensembles even in the presence of large inhomogeneity. We generalize the GBE to include  driving fluctuations and to analyze the coherence under the CCD protocol. By experimentally  measuring the dependence of the coherence time on the second driving strength $\epsilon_m$, we can validate our theoretical analysis and analyze the interplay of competing noise sources. 
In addition to providing a useful tool for protecting known and unknown quantum states, the insights into the CCD dynamics have found applications in high-frequency AC magnetic field sensing \cite{joasQuantumSensingWeak2017,starkNarrowbandwidthSensingHighfrequency2017}. 
The robust driving of the NV center ensemble could further enable the indirect protection of the $^{14}\text{N}$ nuclear spin associated with the NV center, whose coherence is limited by the random telegraph noise caused by the $T_1$ relaxation of the NV electronic spin. The nuclear spin protection requires rapid flips of the NV electron spin \cite{chenProtectingSolidstateSpins2018a}, which can be accomplished by the CCD scheme. Similarly, robust driving of the electronic spin bath~\cite{bauchUltralongDephasingTimes2018}, could enhance the NV coherence time. 
Finally, the scheme demonstrated in this work can be used to design robust quantum control pulses \cite{khanejaUltraBroadbandNMR2016}.  

\section*{Acknowledgments}
This work was supported in part by DARPA DRINQS and NSF PHY1915218. We thank Pai Peng for fruitful discussions and Thanh Nguyen for manuscript revision.

\appendix
\section{Dynamics of the CCD scheme}
To predict the precise dynamics of the CCD scheme, we utilize  Floquet theory to simulate the evolution. The eigenvectors of a time-periodic Hamiltonian are given by $e^{-i\lambda^a t}\Phi^a(t)$ where $\{\lambda^a\}$ are the eigen-energies, and $\Phi^a(t)=\Phi^a(t+T)$ are periodic in time, with $T=\frac{2\pi}{\omega_m}$. 
The evolution of an arbitrary qubit state can then be written as $\Psi(t)=c^+e^{-i\lambda^-t}\Phi^+(t)+c^-e^{-i\lambda^+t}\Phi^-(t)$ with the coefficients $c^\pm$ set by the initial conditions. 
If the initial state is one of the two eigenstates $\Phi^\pm(0)$, then the spin-locking condition is satisfied and the state evolution will only involve one mode, associated with the corresponding $\lambda^\pm$. The  Rabi oscillations will only include frequency components that are integer multiples of $\omega_m$. Otherwise, the evolution will be a superposition of these two modes, and the Rabi oscillations will involve three sets of frequencies, $n\omega_m$ and $n\omega_m\pm (\lambda^+-\lambda^-)$. By tuning the driving parameters, the evolution mode can be well controlled. 

\begin{figure}[htbp]
\centering \includegraphics[width=130mm]{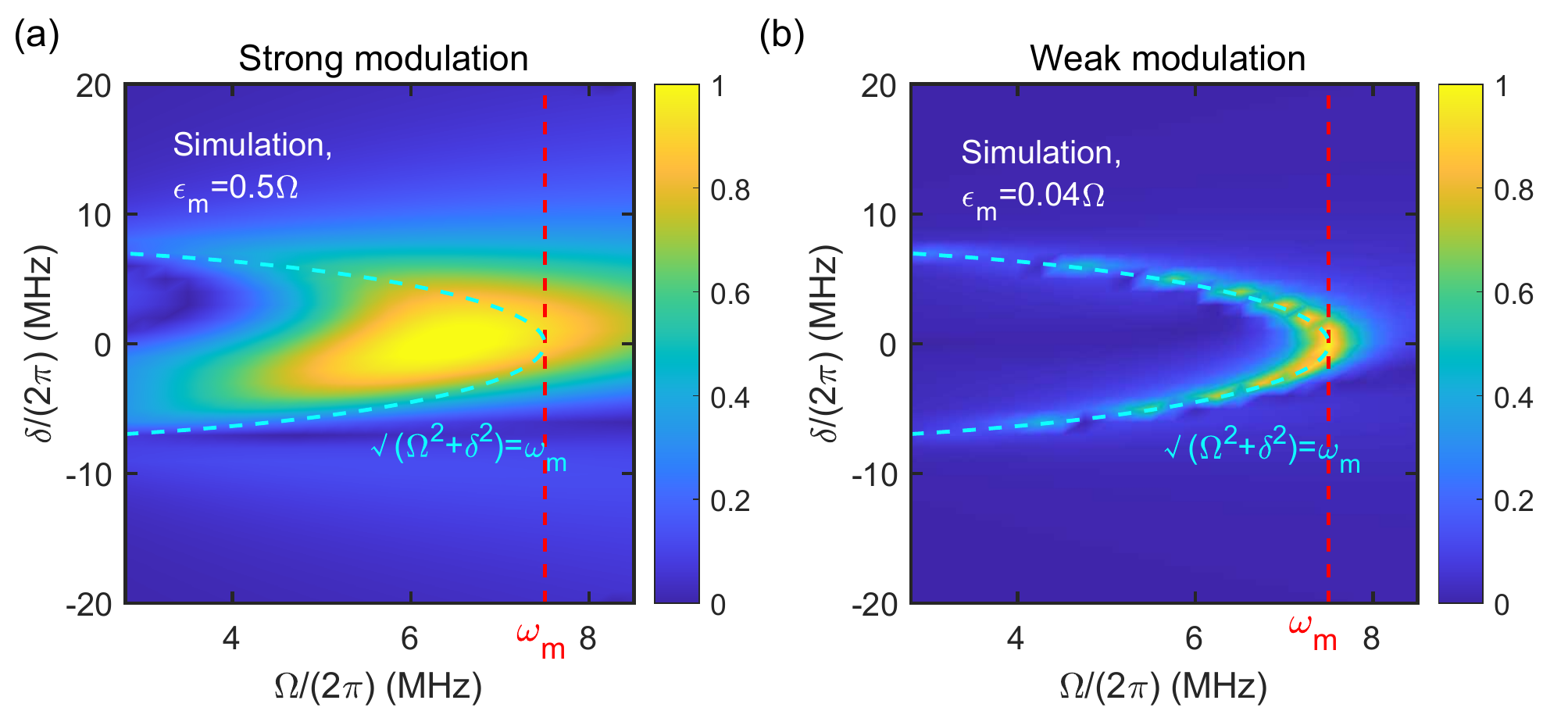}
\caption{
\label{2DRobustness_Appendix} 
Floquet simulation for the experiment in Fig.~\ref{2DRobustness}. The oscillation contrast of the center band is calculated assuming a single NV under the same driving condition as in the experiment seen in Fig.~\ref{2DRobustness}. The decay effects caused by the noise are taken into account by only keeping the initial contrast of the center band while neglecting the contrast contributed from the sidebands in the simulation. The intensity of the colormap represents the contrast value. Under the optimum driving condition, the maximum contrast $(=1)$ is achieved.
Please see Ref.~\cite{wang2020observation} for details on the Floquet simulation.}
\end{figure}

\begin{figure}[htbp]
\centering \includegraphics[width=130mm]{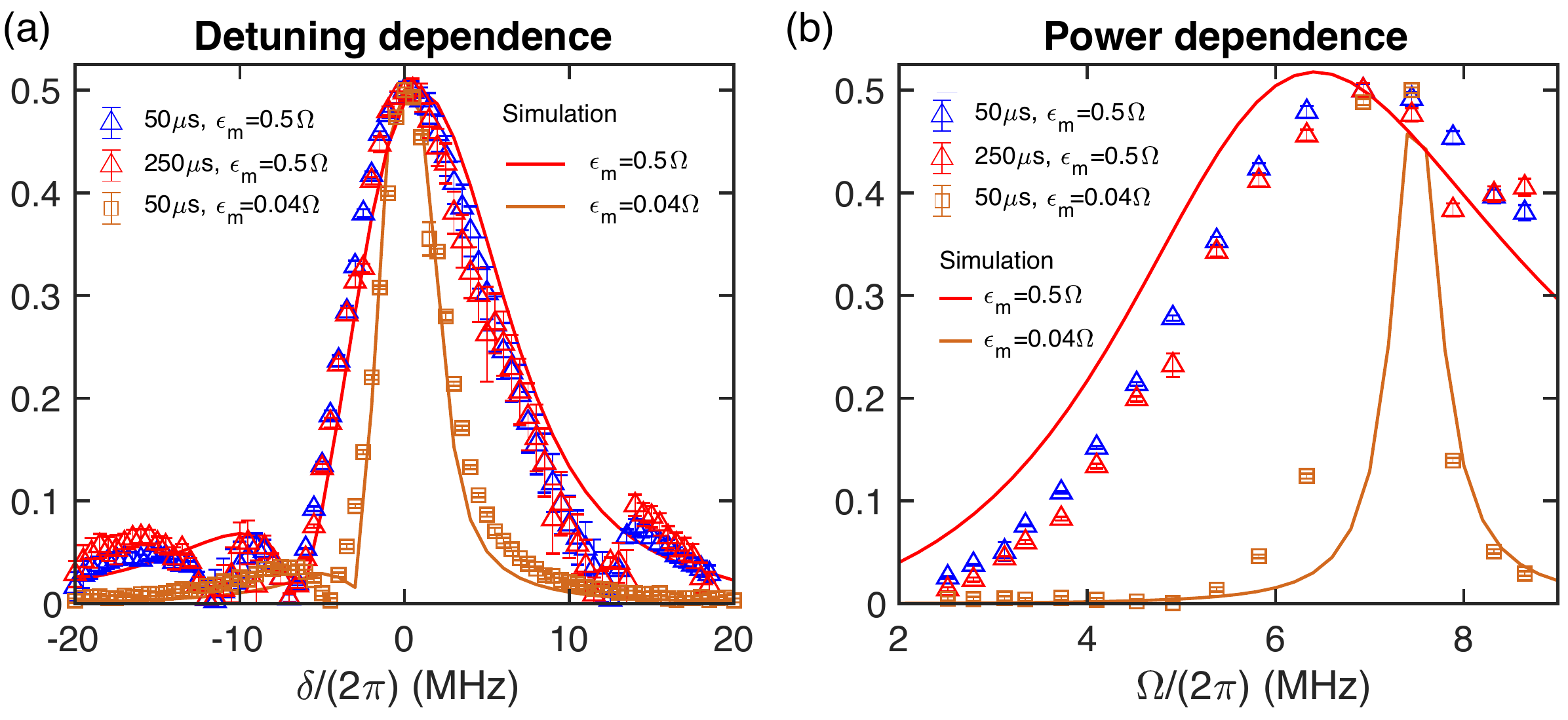}
\caption{\label{2DRobustness_CutPlot} 1D cuts of robustness experiments in Fig.~\ref{2DRobustness} and simulations in Fig.~\ref{2DRobustness_Appendix}. (a) Detuning $\delta$ dependence of center band amplitude, which is a cut of the intensity plot in Fig.~\ref{2DRobustness} (a) along $\Omega=(2\pi)7.5$MHz. Blue triangles and red triangles are data of Rabi oscillation coefficients fitted from Rabi oscillations at time $50\mu $s (from $50\mu $s to $50.5\mu $s) and $250\mu $s (from $250\mu $s to $250.25\mu $s) correspondingly under strong modulation $\epsilon_m=\frac{1}{2}\Omega, \phi=\pi/2$. Yellow squares are data at time $50\mu $s with weak modulation $\epsilon_m=\frac{1}{25}\Omega, \phi=\pi/2$. Red and yellow curves are theoretical prediction of Floquet theory. (b) Power dependence of the oscillation coefficients which is a cut of the intensity plots along $\delta=0$. The colors correspond to those in (a). Note that the plots in (a) and (b) are manually normalized by their maximum value for easier comparison of the peak width, thus the y axis has an arbitrary scale.}
\end{figure}

In the presence of inhomogeneities of the first drive, a stronger modulation is needed to synchronize more ensemble spins to the center band evolution. In the main text (Fig.~\ref{2DRobustness}), we experimentally measured the power and detuning dependence of the center band oscillation contrast (twice the oscillation amplitude) under strong and weak modulation strengths $\epsilon_m$. In Fig.~\ref{2DRobustness_Appendix}, we use Floquet theory to calculate the center band oscillation contrast of a single NV under the experimental conditions, and find a good match with the experiment results. On resonance $\omega_m=\Omega,\delta=0$, the simulation predicts similar contrast for both strong and weak modulation. 
However, in the experiments in Fig.~\ref{2DRobustness} we found that the oscillation contrast was larger with the strong modulation. This can be easily understood by considering that even under a nominal resonance condition, many spins in the ensemble have an offset, due to inhomogeneities, and only a strong drive is enough to achieve a good control. 
In Fig.~\ref{2DRobustness_CutPlot}, we plot 1D cuts of both the experimental data (symbols) and results obtained from simulation (curves). Under strong modulation, the oscillation amplitude as a function of detuning has a  full width at half maximum  much larger than the hyperfine coupling constant $A=2.2$MHz between NV electronic spin and $^{14}$N nuclear spin. This indicates that we can effectively synchronize all nuclear spin sublevels to the center band, and thus protect a known NV state irrespective of the nuclear state. 

\section{Comparison to previous works}
Table~\ref{CoherenceCompare} makes a comparison of our work to previous ones.  An order of magnitude coherence improvement was achieved in both single NV \cite{caiRobustDynamicalDecoupling2012} and sparse NV ensembles \cite{farfurnikExperimentalRealizationTimedependent2017} by applying the CCD scheme with resonant microwave,  and the CCD scheme was also able to improve the coherence of NVs in nano diamonds~\cite{caoProtectingQuantumSpin2020}.  
The CCD  scheme has also been explored by combining high quality mechanical driving, serving as the modulation field, and  microwave driving~\cite{teissierHybridContinuousDynamical2017}; the coherence of a single NV  was improved by  one order of magnitude. 
In comparison, we achieve a 15-fold improvement for the coherence of the two sidebands in a large volume of NV ensembles with $10^{10}$ spins. 
In addition, we also observe a 500-fold improvement for the central band at $\omega_m$, whose long coherence was only previously identified in the mechanical driving experiment. While the latter achieved a similar coherence enhancement, using microwaves only further allows the implementation of mode control of the evolution,  and the phase-modulated CCD scheme can generate larger modulation strength without being limited by microwave power, which provides a more flexible tool for finding optimal coherence and generating even more applications.

\begin{table}[htbp]
\caption{\label{CoherenceCompare} Comparison of Rabi coherence protection with CCD in NV systems. Note that in \cite{teissierHybridContinuousDynamical2017} and this work, coherence times for different frequency components are discussed separately while all the other work only discuss an overall coherence time.}
\begin{tabular}{p{2.2cm}<{\centering}|p{2.2cm}<{\centering}p{2.2cm}<{\centering}p{2.2cm}<{\centering}p{2.2cm}<{\centering}p{2.2cm}<{\centering}}
\hline\hline
\textbf{Parameters} & \textbf{Ref.~\cite{caiRobustDynamicalDecoupling2012} (2012) } & \textbf{Ref.~\cite{farfurnikExperimentalRealizationTimedependent2017} (2017)} & \textbf{Ref.~\cite{teissierHybridContinuousDynamical2017} (2017)} & \textbf{Ref.~\cite{caoProtectingQuantumSpin2020} (2020)} & \textbf{This work} \\ \hline\hline
Sample & Single NV & NV ensemble ($10^4$ spins) & Single NV & Nano diamond & NV ensemble ($10^{10}$ spins) \\\hline
$T_1$ & 1.5ms & 5.9ms & 5.1ms & 0.08735ms & 2.4ms \\ \hline
$\Omega/(2\pi), T_{2\rho}^*$  & 40MHz, 2.3$\mu $s & 9MHz, 0.81$\mu $s & 5.83MHz, 5.3$\mu $s & 8.06MHz & 7.5MHz, 1$\mu $s \\ \hline
$\sigma_\Omega/(2\pi)$ & $\sim 0.1$MHz & $\sim$0.07MHz & $\sim0.2$MHz & $\sim 0.1$MHz & $\sim$0.2MHz \\ \hline
$\epsilon_m/(2\pi)$ & $\sim 1$MHz & $\sim 1$MHz & $\sim 4.1$MHz & $\sim 0.1$MHz & $3$MHz, $1$MHz\\ \hline
$T_{1\rho\rho}$ & \multirow{2}{*}{$21\mu $s} & \multirow{2}{*}{$\sim 14\mu $s} & $\sim$2.9ms & \multirow{2}{*}{$\sim 30\mu $s} & $\sim 0.5$ms \\ \cline{1-1} \cline{4-4} \cline{6-6}
$T_{2\rho\rho}$ &  &  & $\sim 100\mu $s &  & $\sim 15\mu $s \\ \hline\hline
\end{tabular}
\end{table}

\section{Inhomogeneity characterization}
To characterize the inhomogeneity in our sample, we study the power and detuning dependence of the Rabi coherence. 
Assuming that inhomogeneities in the driving power  and static field  are the two main sources of decay for the Rabi oscillations, we simulate the coherence time. We assume a Gaussian distribution of their values,  $f(\Omega+\xi_\Omega,\omega+\xi_\omega)=\frac{1}{2\pi \sigma_\Omega \sigma_{\xi_z}}\exp(-\frac{\xi_\Omega^2}{2\sigma_\Omega^2}-\frac{\xi_\omega^2}{2\sigma_{\omega}^2})$, where $\xi_\Omega$ describes the driving strength inhomogeneity and $\xi_\omega$ the static field along the z axis. 
We take the inhomogeneity of the drive to be proportional to the driving amplitude, $\xi_\Omega=r_\Omega\cdot \Omega$,  where $\xi_\omega$ is fixed. Rabi oscillations are simulated by a two dimensional integration over the power and detuning inhomogeneity
\begin{align*}
    P_{|0\rangle}(t)=&\sum_{i=1}^3 \frac{1}{2} c_i \int_{-\infty}^{\infty}\int_{-\infty}^{\infty} d\xi_\Omega d\xi_\omega f(\Omega+\xi_\Omega,\omega+\xi_\omega) \\
&\frac{\Omega}{\sqrt{\Omega^2+(\omega-\omega_i)^2}} \cos(\sqrt{\Omega^2+(\omega-\omega_i)^2}t)e^{-t/\tau_0}
\end{align*} 
In Fig.~\ref{Noise} of the main text, we plot the simulation results (blue points) when varying the Rabi amplitude and resonance frequency. By comparing the dependence of the simulation results on these parameters under our control with the experiments, we obtain an estimate of the inhomogeneity distribution, $\sigma_\Omega=0.016\Omega$ and $\sigma_{\omega}=(2\pi)0.32\text{MHz}$. 
In simulations, we find that the slope of the $\Omega$-dependence (Fig.~\ref{Noise}a) is more sensitive to the driving inhomogeneity $\sigma_\Omega$, while  the peak width as a function of $\omega$ (Fig.~\ref{Noise}b) is more sensitive to  the static field variance, $\sigma_{\omega}$, which can be explained with the theory in \ref{Appendix_CoherenceLimit}. 
To make the simulation best fit the experiment, we choose an intrinsic coherence time $\tau_0=13\mu $s, which gives a constant offset to the decay rates and may come from other noise sources, such as the spin bath.

\section{Coherence limit}
\label{Appendix_CoherenceLimit}
To analyze the effect of noise on the spin coherence, we consider a semiclassical model, where the noise is taken to be a fluctuating field originating from a classical bath. 
Introducing these stochastic components and assuming $\phi_0=0$,  the amplitude-modulated CCD Hamiltonian reads 
\begin{equation}
    H=\frac{\omega_0}{2}\sigma_z+(\Omega+\xi_{\Omega})\cos(\omega t)\sigma_x-2(\epsilon_m+\xi_{\epsilon_m})\sin(\omega t)\cos(\omega_m t+\phi_m))\sigma_x+\xi_x\sigma_x+\xi_z\sigma_z
\end{equation} 
where $\xi_x,\xi_z$ are the fluctuations of the effective transverse and longitudinal fields and $\xi_{\Omega}$, $\xi_{\epsilon_m}$ are the fluctuations of the driving field. 
Within the RWA, $\Omega\ll\omega$, the Hamiltonian in  the first rotating frame is 
\begin{equation}
    H_I^{(1)} =\bigg[-\frac{\delta}{2}+\xi_z\bigg]\sigma_z+\left[\frac{\Omega+\xi_{\Omega}}{2}+\xi_x\cos(\omega_0 t)\right]\sigma_x+\bigg[(\epsilon_m+\xi_{\epsilon_m})\cos(\omega_m t+\phi_m)-\xi_x\sin(\omega_0 t)\bigg]\sigma_y
\end{equation}
where $\delta=\omega-\omega_0$ is the resonance offset. 
In the following, we will analyze four cases based on this model: on-resonance single driving, off-resonance single driving, amplitude-modulated CCD, and phase-modulated CCD.

\subsection{Single driving with $\delta=0$}
\begin{figure}[h]
\centering \includegraphics[width=100mm]{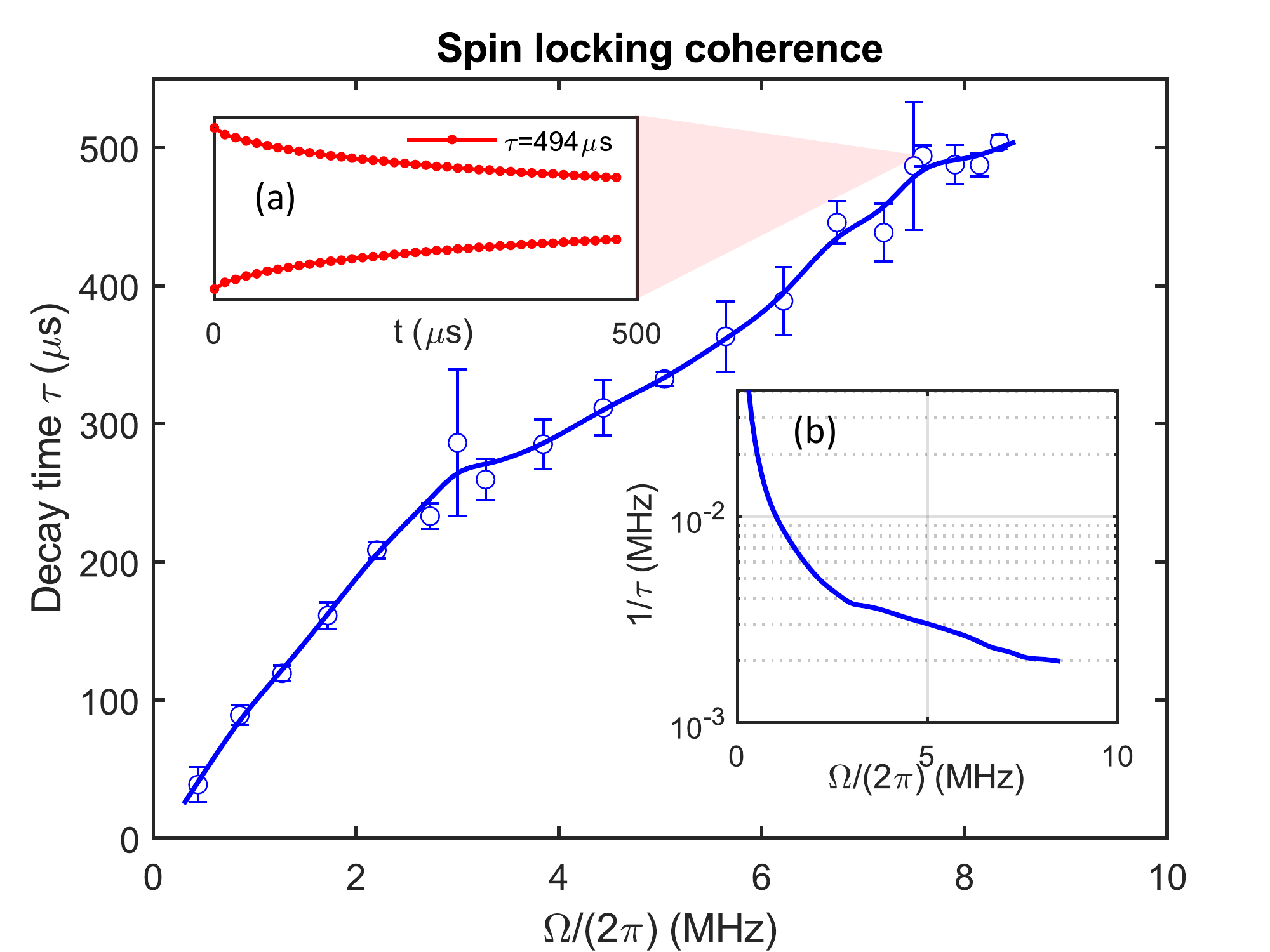}
\caption{\label{SpinLocking}  Spin-locking coherence time $\tau$ as a function of the driving strength $\Omega/(2\pi)$. After initializing the qubit to $\ket{0}$ state, a $\pi/2$ pulse along the y axis rotates the spin to $\frac{1}{\sqrt{2}}(\ket{0}+\ket{1})$, then an on-resonance transverse driving field is applied continuously for a time interval, finally a $\pi/2$ pulse along the y(-y) axis is applied to flip the spin to -z(z) direction for population measurement in $\ket{0}$. By measuring the population in $\ket{0}$ while varying the time intervals between two $\pi/2$ pulses, the spin-locking coherence time can be extracted. Inset (a) is the data plot for $\Omega=(2\pi)7.5\text{MHz}$. Inset (b) is the decay rate $\frac{1}{\tau}$ as a dependence of the driving strength $\Omega/(2\pi)$.
}
\end{figure}

We first analyze the case without the second driving with $\epsilon_m=0$ and compare with previous work in Ref.~\cite{gevaRelaxationTwoLevel1995}. The PSDs in the first rotating frame $S_j^{(1)}$ can be expressed as a function of the PSDs in the lab frame\begin{align}
    S_x^{(1)}(\nu)&=\frac{1}{4}S_{\Omega}(\nu)+\frac{1}{4}\bigg[S_x(\nu+\omega_0)+S_x(\nu-\omega_0)\bigg]\nonumber \\
    S_y^{(1)}(\nu)&=\frac{1}{4}\bigg[S_x(\nu+\omega_0)+S_x(\nu-\omega_0)\bigg]\\
    S_z^{(1)}(\nu)&=S_z(\nu)\nonumber
\end{align}
We can write the decay rates $\Gamma_\alpha$ of the $\alpha=\{x,y,z\}$ components of the qubit Pauli matrix as
\begin{align}
   \Gamma_x &= \frac{1}{4} \bigg[S_x(\omega_0+\Omega)+S_x(\omega_0-\Omega)\bigg]+S_z(\Omega)\nonumber \\
   \Gamma_y &= \frac{1}{2}S_x(\omega_0)+S_z(\Omega)+\frac{1}{4}S_{\Omega}(0)\\
   \Gamma_z &= \frac{1}{2} S_x(\omega_0)+\frac{1}{4}\bigg[S_x(\omega_0+\Omega)+S_x(\omega_0-\Omega)\bigg]+\frac{1}{4}S_{\Omega}(0)\nonumber
\end{align} 
where we used the fact that the decay along one axis is determined by the sum of the rotating frame spectra along the two other axes, $\Gamma_\alpha=S^{(1)}_\beta+S^{(1)}_\gamma$. In turn, these rates can be used to write the longitudinal and transverse relaxation time in the first rotating frame  $T_{1\rho},T_{2\rho}$. With the approximation $S_x(\omega_0\pm\Omega)\approx S_x(\omega_0)$, we obtain
\begin{align}
    \frac{1}{T_{1\rho}}&=\Gamma_x=\frac{1}{2} S_x(\omega_0)+S_z(\Omega)\\
    \frac{1}{T_{2\rho}}&=\frac{1}{2}(\Gamma_y+\Gamma_z)=\frac{3}{4}S_x(\omega_0)+\frac{1}{2}S_z(\Omega)+\frac{1}{4}S_{\Omega}(0)=\frac{1}{2T_{1\rho}}+\frac{1}{T_{2\rho}^{'}}\nonumber
\end{align}
where we defined the pure dephasing time $T_{2\rho^{'}}$ with  $\frac{1}{T_{2\rho}^{'}}=\frac{1}{2}S_x(\omega_0)+\frac{1}{4}S_{\Omega}(0)=\frac{1}{2T_1}+\frac{1}{4}S_{\Omega}(0)$. 
Our analysis up to here is consistent with previous work in Ref.~\cite{gevaRelaxationTwoLevel1995} except for an additional microwave fluctuation term. Fig.~\ref{SpinLocking} is a measurement of spin-locking coherence as a function of $\Omega$. The coherence time increases with $\Omega$ due to the decreasing of $S_z(\Omega)$. The inset (b) plots the decay rate and is a direct measurement of $S_z(\Omega)$~\cite{bylander_noise_2011,yanRotatingframeRelaxationNoise2013}.

\subsection{Single driving with $\delta\neq 0$}
When there is a frequency offset, $\delta\neq 0$, we can diagonalize the non-stochastic Hamiltonian (in the $\sigma_x$ basis) by defining a new set of axes in the first rotating frame with $\sigma_z=\frac{\Omega}{\Omega_R}\sigma_{z^{'}}-\frac{\delta}{\Omega_R}\sigma_{x^{'}},\sigma_x=\frac{\Omega}{\Omega_R}\sigma_{x^{'}}+\frac{\delta}{\Omega_R}\sigma_{z^{'}},\sigma_y=\sigma_{y^{'}}$ where $\Omega_R=\sqrt{\delta^2+\Omega^2}$ is the effective driving field. Then, the Hamiltonian in the first rotating frame becomes
\begin{align}
    H_I^{(1)} =&-\frac{\delta}{2}\sigma_z+ \frac{\Omega}{2}\sigma_x-\xi_x\sin(\omega_0 t)\sigma_y+\left[\frac{\xi_{\Omega}}{2}+\xi_x\cos(\omega_0 t)\right]\sigma_x+\xi_z \sigma_z\\
    =&\frac{\Omega_R}{2}\sigma_{x^{'}}+\xi_z \left[\frac{\Omega}{\Omega_R}\sigma_{z^{'}}-\frac{\delta}{\Omega_R}\sigma_{x^{'}}\right]+\left[\frac{\xi_{\Omega}}{2}+\xi_x\cos(\omega_0 t)\right]\left[\frac{\Omega}{\Omega_R}\sigma_{x^{'}}+\frac{\delta}{\Omega_R}\sigma_{z^{'}}\right]-\xi_x\sin(\omega_0 t)\sigma_{y^{'}}\nonumber
\end{align}
Accordingly, the  PSDs in this modified first rotating frame $S_{\alpha^{'}}^{(1)}$ can be expressed as a function of the PSDs in the lab frame as
\begin{align}
    S_{x^{'}}^{(1)}(\nu)&=\frac{\delta^2}{\Omega_R^2}S_z(\nu)+\frac{\Omega^2}{\Omega_R^2}\left[\frac{1}{4}S_{\Omega}(\nu)+\frac{1}{4}(S_x(\nu+\omega_0)+S_x(\nu-\omega_0)\right]\nonumber\\
    S_{y^{'}}^{(1)}(\nu)&=\frac{1}{4}\bigg[S_x(\nu+\omega_0)+S_x(\nu-\omega_0)\bigg]\\
    S_{z^{'}}^{(1)}(\nu)&=\frac{\Omega^2}{\Omega_R^2}S_z(\nu)+\frac{\delta^2}{\Omega_R^2}\left[\frac{1}{4}S_{\Omega}(\nu)+\frac{1}{4}(S_x(\nu+\omega_0)+S_x(\nu-\omega_0)\right]\nonumber
\end{align}
Similar to what was done above, we can obtain the decay rates of the $\sigma_{\alpha'}$ components and then combine them to obtain the longitudinal and transverse relaxation rates in the rotating frame. 
With the approximation $S_x(\omega_0\pm\Omega_R)\approx S_x(\omega_0)$, we obtain
\begin{align}
    \frac{1}{T_{1\rho}}&=\Gamma_{x^{'}}=\frac{1}{2} S_x(\omega_0)+\frac{\Omega^2}{\Omega_R^2}S_z(\Omega_R)+\frac{\delta^2}{\Omega_R^2}\left[\frac{1}{4}S_{\Omega}(\Omega_R)+\frac{1}{2}S_x(\omega_0)\right]\\
    \frac{1}{T_{2\rho}}&=\frac{1}{2}(\Gamma_{y^{'}}+\Gamma_{z^{'}})\\
&=\frac{\delta^2}{\Omega_R^2}S_z(0)+\frac{\Omega^2}{4\Omega_R^2}\bigg[S_{\Omega}(0)+2S_z(\Omega_R)\bigg]+\frac{1}{8}\frac{\delta^2}{\Omega_R^2}S_\Omega(\Omega_R)+\left[\frac{3}{4}+\frac{\delta^2}{\Omega_R^2}\right]S_x(\omega_0)\nonumber
\end{align}
This analysis  helps explain the experimental results in Fig.~\ref{Noise}(b) where the coherence of Rabi oscillation becomes worse when the detuning increases, due to the $\frac{\delta^2}{\Omega_R^2}S_z(0)$ term in the transverse decay rate $\frac{1}{T_{2\rho}}$. 
For a spin-locking experiment with detuning, instead, the decay rate $\frac{1}{T_{1\rho}}\approx S_x(\omega_0)=\frac{1}{T_1}$ approaches  the $T_1$ relaxation time when $\delta\to\infty$.

\subsection{Amplitude-modulated CCD}
To simplify the calculation, we assume $\omega_0=\omega, \omega_m=\Omega,\phi_0=\phi_m=0$ for all the following discussions. We enter into the second rotating frame defined by $\frac{\omega_m}{2}\sigma_x$ and drop the counter-rotating terms of the driving field but keep the counter-rotating terms of the noise field. The Hamiltonian in the second rotating frame is

\begin{align}
    H_I^{(2)}&=\frac{\epsilon_m}{2}\sigma_y+\left[\frac{\xi_{\Omega}}{2}+\xi_x\cos(\omega_0 t)\right]\sigma_x\nonumber\\
    &+\left[\frac{\xi_{\epsilon_m}}{2}(1+\cos(2\omega_m t))-\xi_x\sin(\omega_0 t)\cos(\omega_m t)+\xi_z\sin(\omega_m t)\right]\sigma_y\\
    &+\left[\frac{\xi_{\epsilon_m}}{2}\sin(2\omega_m t)+\xi_x\sin(\omega_0 t)\sin(\omega_m t)+\xi_z\cos(\omega_m t)\right]\nonumber\sigma_z
\end{align}
The PSDs in the second rotating frame $S_j^{(2)}$ become
\begin{align}
    S_x^{(2)}(\nu)&=\frac{1}{4}S_{\Omega}(\nu)+\frac{1}{4}\bigg[S_x(\nu+\omega_0)+S_x(\nu-\omega_0)\bigg]\nonumber\\
    S_y^{(2)}(\nu)&=\frac{1}{4}S_{\epsilon_m}(\nu)\nonumber\\&+\frac{1}{16}\bigg[S_{\epsilon_m}(\nu+2\omega_m)+S_{\epsilon_m}(\nu-2\omega_m)\bigg]+\frac{1}{4}\bigg[S_z(\nu+\omega_m)+S_z(\nu-\omega_m)\bigg]\\&+\frac{1}{16}\bigg[S_x(\nu+\omega_0+\omega_m)+S_x(\nu+\omega_0-\omega_m)+S_x(\nu-\omega_0+\omega_m)+S_x(\nu-\omega_0-\omega_m)\bigg]\nonumber\\
    S_z^{(2)}(\nu)&=\frac{1}{16}\bigg[S_{\epsilon_m}(\nu+2\omega_m)+S_{\epsilon_m}(\nu-2\omega_m)\bigg]+\frac{1}{4}\bigg[S_z(\nu+\omega_m)+S_z(\nu-\omega_m)\bigg]\nonumber\\&+\frac{1}{16}\bigg[S_x(\nu+\omega_0+\omega_m)+S_x(\nu+\omega_0-\omega_m)+S_x(\nu-\omega_0+\omega_m)+S_x(\nu-\omega_0-\omega_m)\bigg]\nonumber
\end{align}
In the second rotating frame, the static field is along the y axis, and the decay rates can be analyzed in a similar way
\begin{align}
    \Gamma_x&=S_y^{(2)}(0)+S_z^{(2)}(\epsilon_m)\nonumber\\
    \Gamma_y&=S_x^{(2)}(\epsilon_m)+S_z^{(2)}(\epsilon_m)\\
    \Gamma_z&=S_y^{(2)}(0)+S_x^{(2)}(\epsilon_m)\nonumber
\end{align}
Define the longitudinal and transverse relaxation times in the second rotating frame as $T_{1\rho\rho},T_{2\rho\rho}$. Assume that $S_x(\omega_0\pm\Omega\pm\epsilon_m)\approx S_x(\omega_0)$ with $\Omega,\epsilon_m\ll\omega_0$, then
\begin{align}
    \frac{1}{T_{1\rho\rho}}&=\Gamma_y=\frac{1}{4}S_{\Omega}(\epsilon_m)+\frac{3}{4}S_x(\omega_0)\nonumber\\&+\frac{1}{16}\bigg[S_{\epsilon_m}(2\Omega-\epsilon_m)+S_{\epsilon_m}(2\Omega+\epsilon_m)\bigg]+\frac{1}{4}\bigg[S_z(\Omega-\epsilon_m)+S_z(\Omega+\epsilon_m)\bigg] \label{T1rr}\\
    \frac{1}{T_{2\rho\rho}}&=\frac{1}{2}(\Gamma_x+\Gamma_z)=\frac{1}{2T_{1\rho\rho}}+\frac{1}{4} S_{\epsilon_m}(0)+\frac{1}{8} S_{\epsilon_m}(2\Omega)+\frac{1}{2}S_z(\Omega)+\frac{1}{4}S_x(\omega_0)=\frac{1}{T_{2\rho\rho}^{'}}+\frac{1}{2T_{1\rho\rho}} \label{T2rr}
\end{align}
where $\frac{1}{T_{2\rho\rho}^{'}}=\frac{1}{4} S_{\epsilon_m}(0)+\frac{1}{8} S_{\epsilon_m}(2\Omega)+\frac{1}{2}S_z(\Omega)+\frac{1}{4}S_x(\omega_0)$ is defined as the pure dephasing rate in the second rotating frame.

With $\epsilon_m\approx\Omega$ and $S_{\Omega}(\Omega\pm\epsilon_m)\approx S_{\Omega}(\Omega)$, the coherence times in the second rotating frame simplifies to 
\begin{align}
    \frac{1}{T_{1\rho\rho}}&\approx\frac{1}{4} S_{\Omega}(\epsilon_m)+\frac{3}{4}S_x(\omega_0)+\frac{1}{8}S_{\epsilon_m}(2\Omega)+\frac{1}{2}S_z(\Omega)\nonumber\\&=\frac{1}{2T_{1\rho}}+\frac{1}{4} S_{\Omega}(\epsilon_m)+\frac{1}{2}S_x(\omega_0)+\frac{1}{8}S_{\epsilon_m}(2\Omega)\\
    \frac{1}{T_{2\rho\rho}}&\approx\frac{1}{4} S_{\epsilon_m}(0)+\frac{1}{8}S_\Omega(\epsilon_m)+\frac{3}{16} S_{\epsilon_m}(2\Omega)+\frac{3}{4}S_z(\Omega)+\frac{5}{8}S_x(\omega_0)
\end{align}
When $\epsilon_m\approx\Omega$, the approximation here is no longer valid and the coherence is dominated by $S_z(\Omega-\epsilon_m)$.

\subsection{Phase-modulated CCD}
There are two-fold differences in the phase-modulated CCD. First, the modulation amplitude could be assumed in principle to be noise-free, since it arises from the phase modulation that should be very precise, as it depends mostly on the signal source, and not on how it is delivered to the spins. Second, the modulated drive is along the z-direction (instead of the y-direction as it is the case for the amplitude-modulated CCD).
In the lab frame, we assume $\phi_0=0$ and add fluctuation parameters to the phase-modulated CCD Hamiltonian \begin{align}
    H=&\frac{\omega_0}{2}\sigma_z+(\Omega+\xi_{\Omega})\cos(\omega t+2\frac{\epsilon_m}{\Omega}\cos(\omega_m t+\phi_m))\sigma_x+\xi_x\sigma_x+\xi_z\sigma_z
\end{align} 
where $\xi_x,\xi_z$ are the fluctuations of the  transverse and longitudinal fields and $\xi_{\Omega}$ is the fluctuation of the driving field. With the RWA and resonance condition $\Omega\ll\omega=\omega_0$, we can enter into the first rotating frame where
\begin{align}
    H_I^{(1)}& = \frac{\Omega}{2}\sigma_x+\epsilon_m\sin(\omega_m t+\phi_m)\sigma_z+\left[\frac{\xi_{\Omega}}{2}+\xi_x\cos(\omega_0 t+2\frac{\epsilon_m}{\Omega}\cos(\omega_m t+\phi_m))\right]\sigma_x\\
    &-\xi_x\sin(\omega_0 t+2\frac{\epsilon_m}{\Omega}\cos(\omega_m t+\phi_m))\sigma_y+\xi_z \sigma_z\nonumber
\end{align}
Assume $\phi_m=0,\omega_m=\Omega$ and then the Hamiltonian in the second rotating frame is
\begin{align}
    H_I^{(2)}=\frac{\epsilon_m}{2}\sigma_y&+\left[\frac{\xi_{\Omega}}{2}+\xi_x\cos(\omega_0 t+2\frac{\epsilon_m}{\Omega}\cos(\omega_m t))\right]\sigma_x\nonumber\\&
    +\left[\xi_z\sin(\omega_m t)-\xi_x\sin(\omega_0 t+2\frac{\epsilon_m}{\Omega}\cos(\omega_m t))\cos(\omega_m t)\right]\sigma_y\\
    &+\left[xi_z\cos(\omega_m t)+\xi_x\sin(\omega_0 t+2\frac{\epsilon_m}{\Omega}\cos(\omega_m t))\sin(\omega_m t)\right]\sigma_z\nonumber 
\end{align}
The term $\cos(\omega_0 t+2\frac{\epsilon_m}{\Omega}\cos(\omega_m t))$ or $\sin(\omega_0 t+2\frac{\epsilon_m}{\Omega}\cos(\omega_m t))$ can be approximated by calculating the expansion of $\cos(2\frac{\epsilon_m}{\Omega}\cos(\omega_m t))$ or $\sin(2\frac{\epsilon_m}{\Omega}\cos(\omega_m t))$ to first order when $\epsilon_m/\Omega$ is small. For example,
\begin{align}
    \cos(\omega_0 t+2\frac{\epsilon_m}{\Omega}\cos(\omega_m t))&=\cos(\omega_0 t)\cos(2\frac{\epsilon_m}{\Omega}\cos(\omega_m t))-\sin(\omega_0 t)\sin(2\frac{\epsilon_m}{\Omega}\cos(\omega_m t))\nonumber
    \\&\approx \cos(\omega_0 t)-\sin(\omega_0 t)2\frac{\epsilon_m}{\Omega}\cos(\omega_m t)
\end{align}

The PSDs in the second rotating frame $S_j^{(2)}$ become
\begin{align}
    S_x^{(2)}(\nu)&\approx\frac{1}{4}S_\Omega(\nu)+\frac{1}{4}\bigg[S_x(\nu+\omega_0)+S_x(\nu-\omega_0)\bigg]\nonumber\\&+\frac{1}{4}(\frac{\epsilon_m}{\Omega})^2\bigg[S_x(\nu+\omega_0+\omega_m)+S_x(\nu+\omega_0-\omega_m)+S_x(\nu-\omega_0+\omega_m)+S_x(\nu-\omega_0-\omega_m)\bigg]\nonumber\\
    S_y^{(2)}(\nu)&\approx \frac{1}{4}\bigg[S_z(\nu-\omega_m)+S_z(\nu+\omega_m)\bigg]\nonumber\\&+\frac{1}{16}\bigg[S_x(\nu+\omega_0+\omega_m)+S_x(\nu+\omega_0-\omega_m)+S_x(\nu-\omega_0+\omega_m)+S_x(\nu-\omega_0-\omega_m)\bigg]\nonumber\\&+\frac{1}{4}(\frac{\epsilon_m}{\Omega})^2\bigg[S_x(\nu+\omega_0)+S_x(\nu-\omega_0)\bigg]\\&+\frac{1}{16}(\frac{\epsilon_m}{\Omega})^2\bigg[S_x(\nu+\omega_0+2\omega_m)+S_x(\nu+\omega_0-2\omega_m)+S_x(\nu-\omega_0+2\omega_m)+S_x(\nu-\omega_0-2\omega_m)\bigg]\nonumber\\
    S_z^{(2)}(\nu)&\approx \frac{1}{4}\bigg[S_z(\nu-\omega_m)+S_z(\nu+\omega_m)\bigg]\nonumber\\&+\frac{1}{16}\bigg[S_x(\nu+\omega_0+\omega_m)+S_x(\nu+\omega_0-\omega_m)+S_x(\nu-\omega_0+\omega_m)+S_x(\nu-\omega_0-\omega_m)\bigg]\nonumber\\&+\frac{1}{16}(\frac{\epsilon_m}{\Omega})^2\bigg[S_x(\nu+\omega_0+2\omega_m)+S_x(\nu+\omega_0-2\omega_m)+S_x(\nu-\omega_0+2\omega_m)+S_x(\nu-\omega_0-2\omega_m)\bigg]\nonumber
\end{align}
In the second rotating frame, the static field is along the y axis, the decay rates can be analyzed in a similar way
\begin{align}
    \Gamma_x&=S_y^{(2)}(0)+S_z^{(2)}(\epsilon_m)\nonumber\\
    \Gamma_y&=S_x^{(2)}(\epsilon_m)+S_z^{(2)}(\epsilon_m)\\
    \Gamma_z&=S_y^{(2)}(0)+S_x^{(2)}(\epsilon_m)\nonumber
\end{align}
The longitudinal and transverse relaxation time  $T_{1\rho\rho},T_{2\rho\rho}$, under resonance condition $\Omega=\omega_m$ and assuming  $S_x(\omega_0\pm\Omega\pm\epsilon_m)\approx S_x(\omega_0)$ with $\Omega,\epsilon_m\ll\omega_0$, are given by
\begin{align}
    \frac{1}{T_{1\rho\rho}}&=\Gamma_y=\bigg[\frac{3}{4}+\frac{5}{4}(\frac{\epsilon_m}{\Omega})^2\bigg]S_x(\omega_0)+\frac{1}{4}S_\Omega(\epsilon_m)+\frac{1}{4}\bigg[S_z(\Omega-\epsilon_m)+S_z(\Omega+\epsilon_m)\bigg] \\
    \frac{1}{T_{2\rho\rho}}&=\frac{1}{2}(\Gamma_x+\Gamma_z)=\frac{1}{2T_{1\rho\rho}}+\frac{1}{2}S_z(\Omega)+\bigg[\frac{1}{4}+\frac{3}{4}(\frac{\epsilon_m}{\Omega})^2\bigg]S_x(\omega_0)=\frac{1}{T_{2\rho\rho}^{'}}+\frac{1}{2T_{1\rho\rho}} 
\end{align}
where $\frac{1}{T_{2\rho\rho}^{'}}=\frac{1}{2}S_z(\Omega)+(\frac{1}{4}+\frac{3}{4}(\frac{\epsilon_m}{\Omega})^2)S_x(\omega_0)$ is defined as the pure dephasing rate in the second rotating frame.

With $\epsilon_m\ll\Omega$ and $S_{\Omega}(\Omega\pm\epsilon_m)\approx S_{\Omega}(\Omega)$, the longitudinal coherence time in the second rotating frame becomes \begin{align}
    \frac{1}{T_{1\rho\rho}}\approx&\frac{1}{4} S_{\Omega}(\epsilon_m)+\frac{3}{4}S_x(\omega_0)+\frac{1}{2}S_z(\Omega)=\frac{1}{2T_{1\rho}}+\frac{1}{4} S_{\Omega}(\epsilon_m)+\frac{1}{2}S_x(\omega_0)
\end{align} The transverse coherence time in the second rotating frame \begin{align}
    \frac{1}{T_{2\rho\rho}}\approx&\frac{1}{8}S_\Omega(\epsilon_m)+\frac{3}{4}S_z(\Omega)+\frac{5}{8}S_x(\omega_0)
\end{align} Expressions here can also be obtained by simply setting $\xi_{\epsilon_m}=0$ in the amplitude-modulated situation.


\section*{References}
\bibliographystyle{unsrt}

\bibliography{ModulatedDriving}

\begin{thebibliography}{10}

\bibitem{hahnSpinEchoes1950}
E.~L. Hahn.
\newblock Spin {{Echoes}}.
\newblock {\em Physical Review}, 80(4):580--594, November 1950.

\bibitem{carrEffectsDiffusionFree1954}
H.~Y. Carr and E.~M. Purcell.
\newblock Effects of {{Diffusion}} on {{Free Precession}} in {{Nuclear Magnetic
  Resonance Experiments}}.
\newblock {\em Physical Review}, 94(3):630--638, May 1954.

\bibitem{meiboomModifiedSpinEcho1958}
S.~Meiboom and D.~Gill.
\newblock Modified {{Spin}}-{{Echo Method}} for {{Measuring Nuclear Relaxation
  Times}}.
\newblock {\em Review of Scientific Instruments}, 29(8):688--691, August 1958.

\bibitem{ryanRobustDecouplingTechniques2010a}
C.~A. Ryan, J.~S. Hodges, and D.~G. Cory.
\newblock Robust {{Decoupling Techniques}} to {{Extend Quantum Coherence}} in
  {{Diamond}}.
\newblock {\em Physical Review Letters}, 105(20):200402, November 2010.

\bibitem{souzaRobustDynamicalDecoupling2012}
Alexandre~M. Souza, Gonzalo~A. {\'A}lvarez, and Dieter Suter.
\newblock Robust dynamical decoupling.
\newblock {\em Philosophical Transactions of the Royal Society A: Mathematical,
  Physical and Engineering Sciences}, 370(1976):4748--4769, October 2012.

\bibitem{uhrigKeepingQuantumBit2007}
G{\"o}tz~S. Uhrig.
\newblock Keeping a {{Quantum Bit Alive}} by {{Optimized}} {$\pi$} -{{Pulse
  Sequences}}.
\newblock {\em Physical Review Letters}, 98(10):100504, March 2007.

\bibitem{yangUniversalityUhrigDynamical2008}
Wen Yang and Ren-Bao Liu.
\newblock Universality of {{Uhrig Dynamical Decoupling}} for {{Suppressing
  Qubit Pure Dephasing}} and {{Relaxation}}.
\newblock {\em Physical Review Letters}, 101(18):180403, October 2008.

\bibitem{biercukOptimizedDynamicalDecoupling2009}
Michael~J. Biercuk, Hermann Uys, Aaron~P. VanDevender, Nobuyasu Shiga, Wayne~M.
  Itano, and John~J. Bollinger.
\newblock Optimized dynamical decoupling in a model quantum memory.
\newblock {\em Nature}, 458(7241):996--1000, April 2009.

\bibitem{mukhtarUniversalDynamicalDecoupling2010}
Musawwadah Mukhtar, Thuan~Beng Saw, Wee~Tee Soh, and Jiangbin Gong.
\newblock Universal dynamical decoupling: {{Two}}-qubit states and beyond.
\newblock {\em Physical Review A}, 81(1):012331, January 2010.

\bibitem{violaRobustDynamicalDecoupling2003}
Lorenza Viola and Emanuel Knill.
\newblock Robust {{Dynamical Decoupling}} of {{Quantum Systems}} with {{Bounded
  Controls}}.
\newblock {\em Physical Review Letters}, 90(3):037901, January 2003.

\bibitem{gordonOptimalDynamicalDecoherence2008}
Goren Gordon, Gershon Kurizki, and Daniel~A. Lidar.
\newblock Optimal {{Dynamical Decoherence Control}} of a {{Qubit}}.
\newblock {\em Physical Review Letters}, 101(1):010403, July 2008.

\bibitem{fanchiniContinuouslyDecouplingSinglequbit2007}
F.~F. Fanchini, J.~E.~M. Hornos, and R.~d.~J. Napolitano.
\newblock Continuously decoupling single-qubit operations from a perturbing
  thermal bath of scalar bosons.
\newblock {\em Physical Review A}, 75(2):022329, February 2007.

\bibitem{hiroseContinuousDynamicalDecoupling2012}
Masashi Hirose, Clarice~D. Aiello, and Paola Cappellaro.
\newblock Continuous dynamical decoupling magnetometry.
\newblock {\em Physical Review A}, 86(6):062320, December 2012.

\bibitem{laraouiRotatingFrameSpin2011}
Abdelghani Laraoui and Carlos~A. Meriles.
\newblock Rotating frame spin dynamics of a nitrogen-vacancy center in a
  diamond nanocrystal.
\newblock {\em Physical Review B}, 84(16):161403, October 2011.

\bibitem{mkhitaryanDecayRotaryEchoes2014}
V.~V. Mkhitaryan and V.~V. Dobrovitski.
\newblock Decay of the rotary echoes for the spin of a nitrogen-vacancy center
  in diamond.
\newblock {\em Physical Review B}, 89(22):224402, June 2014.

\bibitem{aharonFullyRobustQubit2016}
N~Aharon, I~Cohen, F~Jelezko, and A~Retzker.
\newblock Fully robust qubit in atomic and molecular three-level systems.
\newblock {\em New Journal of Physics}, 18(12):123012, December 2016.

\bibitem{starkClockTransitionContinuous2018}
Alexander Stark, Nati Aharon, Alexander Huck, Haitham A.~R. {El-Ella}, Alex
  Retzker, Fedor Jelezko, and Ulrik~L. Andersen.
\newblock Clock transition by continuous dynamical decoupling of a three-level
  system.
\newblock {\em Scientific Reports}, 8(1):14807, December 2018.

\bibitem{loretzRadioFrequencyMagnetometryUsing2013}
M.~Loretz, T.~Rosskopf, and C.~L. Degen.
\newblock Radio-{{Frequency Magnetometry Using}} a {{Single Electron Spin}}.
\newblock {\em Physical Review Letters}, 110(1):017602, January 2013.

\bibitem{DuCoherenceProtectedPhysRevLett.109.070502}
Xiangkun Xu, Zixiang Wang, Changkui Duan, Pu~Huang, Pengfei Wang, Ya~Wang,
  Nanyang Xu, Xi~Kong, Fazhan Shi, Xing Rong, and Jiangfeng Du.
\newblock Coherence-protected quantum gate by continuous dynamical decoupling
  in diamond.
\newblock {\em Phys. Rev. Lett.}, 109:070502, Aug 2012.

\bibitem{caiRobustDynamicalDecoupling2012}
J-M Cai, B~Naydenov, R~Pfeiffer, L~P McGuinness, K~D Jahnke, F~Jelezko, M~B
  Plenio, and A~Retzker.
\newblock Robust dynamical decoupling with concatenated continuous driving.
\newblock {\em New Journal of Physics}, 14(11):113023, November 2012.

\bibitem{khanejaUltraBroadbandNMR2016}
Navin Khaneja, Abhinav Dubey, and Hanudatta~S. Atreya.
\newblock Ultra broadband {{NMR}} spectroscopy using multiple rotating frame
  technique.
\newblock {\em Journal of Magnetic Resonance}, 265:117--128, April 2016.

\bibitem{saikoSuppressionElectronSpin2018}
A.P. Saiko, R.~Fedaruk, and S.A. Markevich.
\newblock Suppression of electron spin decoherence in {{Rabi}} oscillations
  induced by an inhomogeneous microwave field.
\newblock {\em Journal of Magnetic Resonance}, 290:60--67, May 2018.

\bibitem{cohenContinuousDynamicalDecoupling2017}
I.~Cohen, N.~Aharon, and A.~Retzker.
\newblock Continuous dynamical decoupling utilizing time-dependent detuning:
  {{Continuous}} dynamical decoupling utilizing time-dependent detuning.
\newblock {\em Fortschritte der Physik}, 65(6-8):1600071, June 2017.

\bibitem{farfurnikExperimentalRealizationTimedependent2017}
D.~Farfurnik, N.~Aharon, I.~Cohen, Y.~Hovav, A.~Retzker, and N.~{Bar-Gill}.
\newblock Experimental realization of time-dependent phase-modulated continuous
  dynamical decoupling.
\newblock {\em Physical Review A}, 96(1):013850, July 2017.

\bibitem{rohrSynchronizingDynamicsSingle2014}
S.~Rohr, E.~{Dupont-Ferrier}, B.~Pigeau, P.~Verlot, V.~Jacques, and O.~Arcizet.
\newblock Synchronizing the {{Dynamics}} of a {{Single Nitrogen Vacancy Spin
  Qubit}} on a {{Parametrically Coupled Radio}}-{{Frequency Field}} through
  {{Microwave Dressing}}.
\newblock {\em Physical Review Letters}, 112(1):010502, January 2014.

\bibitem{laytonRabiResonanceSpin2014}
Kelvin~J. Layton, Bahman Tahayori, Iven~M.Y. Mareels, Peter~M. Farrell, and
  Leigh~A. Johnston.
\newblock Rabi resonance in spin systems: {{Theory}} and experiment.
\newblock {\em Journal of Magnetic Resonance}, 242:136--142, May 2014.

\bibitem{saikoMultiphotonTransitionsRabi2015}
Alexander~P. Saiko, Ryhor Fedaruk, and Siarhei~A. Markevich.
\newblock Multi-photon transitions and {{Rabi}} resonance in continuous wave
  {{EPR}}.
\newblock {\em Journal of Magnetic Resonance}, 259:47--55, October 2015.

\bibitem{teissierHybridContinuousDynamical2017}
Jean Teissier, Arne Barfuss, and Patrick Maletinsky.
\newblock Hybrid continuous dynamical decoupling: A photon-phonon doubly
  dressed spin.
\newblock {\em Journal of Optics}, 19(4):044003, April 2017.

\bibitem{bertainaExperimentalProtectionQubit2020}
Sylvain Bertaina, Herv{\'e} Vezin, and Irinel Chiorescu.
\newblock Experimental protection of qubit coherence by using a phase-tunable
  image drive.
\newblock {\em arXiv:2001.02417 [cond-mat, physics:quant-ph]}, January 2020.

\bibitem{caoProtectingQuantumSpin2020}
Q.-Y. Cao, P.-C. Yang, M.-S. Gong, M.~Yu, A.~Retzker, M.B. Plenio,
  C.~M{\"u}ller, N.~Tomek, B.~Naydenov, L.P. McGuinness, F.~Jelezko, and J.-M.
  Cai.
\newblock Protecting {{Quantum Spin Coherence}} of {{Nanodiamonds}} in {{Living
  Cells}}.
\newblock {\em Physical Review Applied}, 13(2):024021, February 2020.

\bibitem{wang2020observation}
Guoqing Wang, Yi-Xiang Liu, and Paola Cappellaro.
\newblock Observation of high-order {{Mollow}} triplet by quantum mode control
  with concatenated continuous driving.
\newblock {\em arXiv:2008.06435 [physics, physics:quant-ph]}, August 2020.

\bibitem{dohertyNitrogenvacancyColourCentre2013a}
Marcus~W. Doherty, Neil~B. Manson, Paul Delaney, Fedor Jelezko, J{\"o}rg
  Wrachtrup, and Lloyd~C.L. Hollenberg.
\newblock The nitrogen-vacancy colour centre in diamond.
\newblock {\em Physics Reports}, 528(1):1--45, July 2013.

\bibitem{phamEnhancedSolidstateMultispin2012}
L.~M. Pham, N.~{Bar-Gill}, C.~Belthangady, D.~Le~Sage, P.~Cappellaro, M.~D.
  Lukin, A.~Yacoby, and R.~L. Walsworth.
\newblock Enhanced solid-state multispin metrology using dynamical decoupling.
\newblock {\em Physical Review B}, 86(4):045214, July 2012.

\bibitem{bauchUltralongDephasingTimes2018}
Erik Bauch, Connor~A. Hart, Jennifer~M. Schloss, Matthew~J. Turner, John~F.
  Barry, Pauli Kehayias, Swati Singh, and Ronald~L. Walsworth.
\newblock Ultralong {{Dephasing Times}} in {{Solid}}-{{State Spin Ensembles}}
  via {{Quantum Control}}.
\newblock {\em Physical Review X}, 8(3):031025, July 2018.

\bibitem{naydenovDynamicalDecouplingSingleelectron2011}
Boris Naydenov, Florian Dolde, Liam~T. Hall, Chang Shin, Helmut Fedder, Lloyd
  C.~L. Hollenberg, Fedor Jelezko, and J{\"o}rg Wrachtrup.
\newblock Dynamical decoupling of a single-electron spin at room temperature.
\newblock {\em Physical Review B}, 83(8):081201, February 2011.

\bibitem{shimRobustDynamicalDecoupling2012}
J.~H. Shim, I.~Niemeyer, J.~Zhang, and D.~Suter.
\newblock Robust dynamical decoupling for arbitrary quantum states of a single
  {{NV}} center in diamond.
\newblock {\em EPL (Europhysics Letters)}, 99(4):40004, August 2012.

\bibitem{bar-gillSolidstateElectronicSpin2013}
N.~{Bar-Gill}, L.M. Pham, A.~Jarmola, D.~Budker, and R.L. Walsworth.
\newblock Solid-state electronic spin coherence time approaching one second.
\newblock {\em Nature Communications}, 4(1):1743, June 2013.

\bibitem{jaskulaPhysRevApplied.11.054010}
J.-C. Jaskula, K.~Saha, A.~Ajoy, D.J. Twitchen, M.~Markham, and P.~Cappellaro.
\newblock Cross-sensor feedback stabilization of an emulated quantum spin
  gyroscope.
\newblock {\em Phys. Rev. Applied}, 11:054010, May 2019.

\bibitem{gevaRelaxationTwoLevel1995}
Eitan Geva, Ronnie Kosloff, and J.~L. Skinner.
\newblock On the relaxation of a two-level system driven by a strong
  electromagnetic field.
\newblock {\em The Journal of Chemical Physics}, 102(21):8541--8561, June 1995.

\bibitem{jingDecoherenceElectricallyDriven2014}
Jun Jing, Peihao Huang, and Xuedong Hu.
\newblock Decoherence of an electrically driven spin qubit.
\newblock {\em Physical Review A}, 90(2):022118, August 2014.

\bibitem{yanRotatingframeRelaxationNoise2013}
Fei Yan, Simon Gustavsson, Jonas Bylander, Xiaoyue Jin, Fumiki Yoshihara,
  David~G. Cory, Yasunobu Nakamura, Terry~P. Orlando, and William~D. Oliver.
\newblock Rotating-frame relaxation as a noise spectrum analyser of a
  superconducting qubit undergoing driven evolution.
\newblock {\em Nature Communications}, 4(1):2337, December 2013.

\bibitem{joasQuantumSensingWeak2017}
T.~Joas, A.~M. Waeber, G.~Braunbeck, and F.~Reinhard.
\newblock Quantum sensing of weak radio-frequency signals by pulsed {{Mollow}}
  absorption spectroscopy.
\newblock {\em Nature Communications}, 8(1):964, December 2017.

\bibitem{starkNarrowbandwidthSensingHighfrequency2017}
Alexander Stark, Nati Aharon, Thomas Unden, Daniel Louzon, Alexander Huck, Alex
  Retzker, Ulrik~L. Andersen, and Fedor Jelezko.
\newblock Narrow-bandwidth sensing of high-frequency fields with continuous
  dynamical decoupling.
\newblock {\em Nature Communications}, 8(1):1105, December 2017.

\bibitem{chenProtectingSolidstateSpins2018a}
Mo~Chen, Won Kyu~Calvin Sun, Kasturi Saha, Jean-Christophe Jaskula, and Paola
  Cappellaro.
\newblock Protecting solid-state spins from a strongly coupled environment.
\newblock {\em New Journal of Physics}, 20(6):063011, June 2018.

\bibitem{bylander_noise_2011}
Jonas Bylander, Simon Gustavsson, Fei Yan, Fumiki Yoshihara, Khalil Harrabi,
  George Fitch, David~G. Cory, Yasunobu Nakamura, Jaw-Shen Tsai, and William~D.
  Oliver.
\newblock Noise spectroscopy through dynamical decoupling with a
  superconducting flux qubit.
\newblock {\em Nature Physics}, 7(7):565--570, July 2011.

\end{thebibliography}

\end{document}